\documentclass[a4paper,fleqn]{cas-sc}
\usepackage{placeins}
\usepackage{setspace}
\usepackage{graphicx}
\usepackage{float}
\usepackage{subcaption}
\usepackage{hyperref}  
\usepackage{amsmath}
\usepackage{booktabs}
\usepackage{array}
\usepackage[hypcap=false]{caption}
\usepackage{multicol}
\usepackage{gensymb}
\usepackage{tabularx}
\usepackage{ragged2e}
\usepackage[version=4]{mhchem}

\usepackage[authoryear,longnamesfirst,sort&compress]{natbib}
\usepackage{anyfontsize}
\def\tsc#1{\csdef{#1}{\textsc{\lowercase{#1}}\xspace}}
\tsc{WGM}
\tsc{QE}

\makeatletter
\renewcommand{\fps@figure}{htbp}
\renewcommand{\fps@table}{htbp}
\makeatother

\begin{document}
\let\WriteBookmarks\relax

\shorttitle{}    

\shortauthors{Shandilaya, Alaleeli, Kim, Mobasher, and Roshankhah}  

\title [mode = title]{Local Stress Redistribution Controls Interactions between Hydraulic Fractures and Pre-existing Fractures}  

%

\author{S. Shandilaya}[orcid=0009-0006-1582-8574]
\fnmark[a, $\dagger$]

\credit{Conducting hydraulic fracturing tests, specimen preparation, DIC analyses, manuscript writing, and editing}
\affiliation[1]{organization={Civil and Environmental Engineering, University of Utah},
            city={Salt Lake City},
            state={Utah},
            country={USA}}
\author{M. Alaleeli}[orcid=0000-0001-9560-2036]
\fnmark[b, $\dagger$]
\credit{XFEM modeling framework, numerical simulations and analysis, manuscript writing and editing}
\author{S.H. Kim}[orcid=0009-0005-7444-248X]
\fnmark[a]
\credit{Conducting hydraulic fracturing tests, DIC analyses, appendix-A writing and editing}
\author{M. Mobasher}
\fnmark[b]
\affiliation[2]{organization={Department of Civil and Urban Engineering, New York University},
            city={Abu Dhabi},
            country={United Arab Emirates}}
\credit{supervising XFEM modeling, manuscript writing, and editing}

\author{S. Roshankhah}[orcid=0000-0002-1160-7882]
\fnmark[a]
\cormark[1]
\ead{Shahrzad.Roshankhah@utah.edu}
\ead[url]{https://sites.google.com/view/geoserl/home}
\credit{Conceptualization of this study, supervising the laboratory experimental study and analyses, manuscript writing, and editing}

\cortext[1]{Corresponding author}

\nonumnote{$\dagger$ These authors contributed equally to this work and share first authorship.}

\begin{abstract}
Hydraulic fracture (HF) propagation in naturally fractured formations is strongly influenced by local stress states in the vicinity of pre-existing natural fractures (NFs). While NFs are known to perturb stresses, the role of NF-induced shear deformation and stress redistribution in controlling HF trajectories remains poorly characterized. This study investigates how NF-induced stress redistribution governs HF-NF interactions through coupled laboratory experiments and poroelastic extended finite element simulations. Water-injection tests are performed on intact and pre-fractured PMMA specimens under plane-strain conditions. Digital image correlation provides full-field measurements of displacement and strain evolution during mechanical loading and hydraulic fracturing. Under fixed-base, lateral confinement, and vertical compression boundary conditions, inclined NFs induce asymmetric stress redistribution and shear deformation, generating distinct local stress states prior to fluid injection. The results demonstrate that HF trajectory is governed by the sign and spatial distribution of shear stress and shear strain components generated by NF orientation relative to the far-field maximum principal stress. Shear deformation that promotes compressive stress development adjacent to the NF causes the HF to deflect away, whereas shear deformation that reduces the effective normal stress along the NF promotes fracture attraction and linkage. Corresponding numerical reproduction of HF curvature in pre-fractured specimens requires mixed-mode (Mode~I–II) fracture energy release criteria, while the intact specimen propagates in pure Mode~I. Overall, the findings reveal a transition from tensile opening to shear-assisted mixed-mode propagation as local stress states evolve due to the presence of NFs. These insights provide a mechanistic basis for predicting and controlling fracture trajectories in subsurface stimulation and storage applications.
\end{abstract}




\maketitle

\makeatletter
\setlength{\@fptop}{0pt}
\setlength{\@fpsep}{8pt}
\setlength{\@fpbot}{0pt plus 1fil}
\renewcommand{\fps@figure}{htbp}
\renewcommand{\fps@table}{htbp}
\makeatother


\section{Introduction}
Hydraulic fracturing enhances fluid transport in low-permeability formations by creating new flow pathways through high-rate fluid injection. Since its first field application in the 1940s (Montgomery \& Smith, \citeyear{montgomery2010hydraulic}), it has become central to subsurface resource management, supporting shale gas and tight oil production, geothermal energy extraction, mining preconditioning, as well as geological storage technologies such as \ce{CO2} sequestration (Adachi et al., \citeyear{adachi2007computer}; King, \citeyear{king2010thirty}; Moska et al., \citeyear{moska2021hydraulic}; Chen et al., \citeyear{chen2022review}; Xie et al., \citeyear{xie2024state}). The objectives of subsurface fluid injection vary across applications. Resource recovery operations require complex fracture networks that maximize reservoir contact and fluid transport, whereas storage formations demand minimal uncontrolled fracture propagation to ensure sealing integrity and long-term safety. In both cases, understanding fracture initiation and propagation is essential for optimizing performance and mitigating risks such as leakage or induced seismicity (King, \citeyear{king2010thirty}; Chen et al., \citeyear{chen2024caprock}).

\begin{figure}[]
  \centering
  \begin{minipage}[b]{0.45\textwidth}
    \centering
    \begin{subfigure}[b]{0.44\linewidth}
      \includegraphics[width=\linewidth]{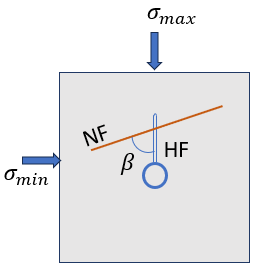}
      \caption{Cross}
      \label{fig:crossing}
    \end{subfigure}
    \hfill
    \begin{subfigure}[b]{0.44\linewidth}
      \includegraphics[width=\linewidth]{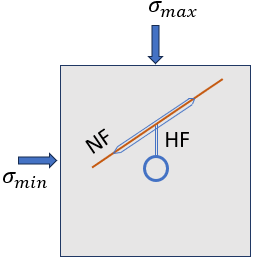}
      \caption{Dilate}
      \label{fig:Penetration}
    \end{subfigure}
    \vspace{1em}
    \begin{subfigure}[b]{0.44\linewidth}
      \includegraphics[width=\linewidth]{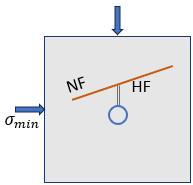}
      \caption{Arrest}
      \label{fig:arrest}
    \end{subfigure}
    \hfill
    \begin{subfigure}[b]{0.44\linewidth}
      \includegraphics[width=\linewidth]{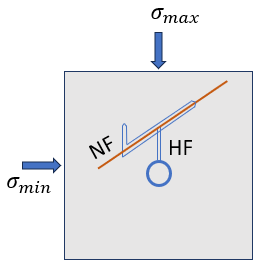}
      \caption{Offset}
      \label{fig:Offset}
    \end{subfigure}
  \end{minipage}%
  \hfill
  \begin{minipage}[b]{0.50\textwidth}
    \centering
    \begin{subfigure}[b]{\linewidth}
      \includegraphics[height=7cm]{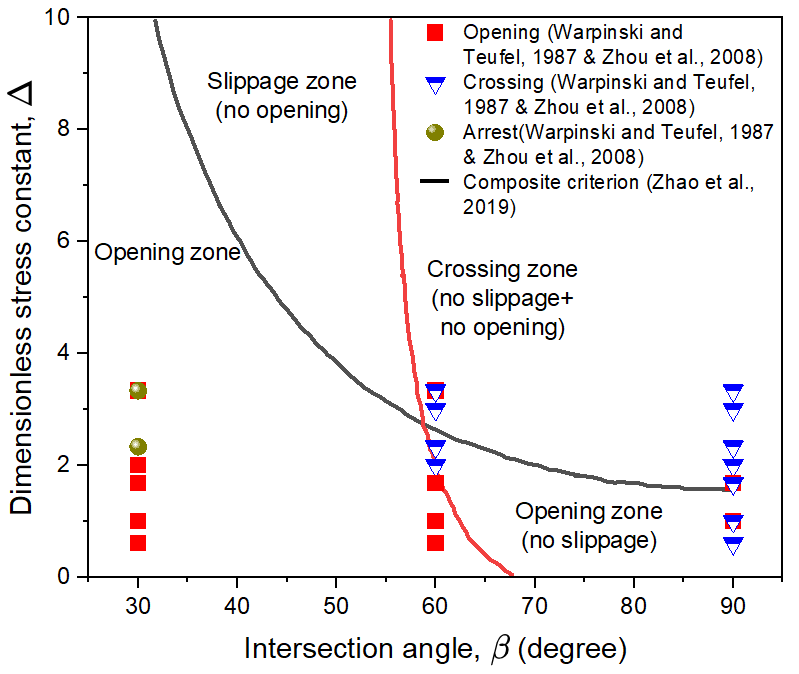}
      \caption{HF–NF interaction at different intersection angles ($\beta$)}
      \label{fig:zone}
    \end{subfigure}
  \end{minipage}
  \caption{Schematics of different interaction scenarios between a natural fracture (NF) and a hydraulic fracture (HF), under anisotropic compressive stress conditions with $\sigma_{\max}$ in the vertical direction. (a) HF crosses NF, (b) HF dilates NF, (c) HF arrests at NF, (d) HF offsets along NF, and (e) interaction regime map outlining conditions for crossing, slippage, arrest, or offset based on the dimensionless stress constant $\Delta = \frac{\sigma_H - \sigma_h}{\sigma_h}$ (after Chen et al., \citeyear{chen2022review}; Zhao et al., \citeyear{zhao2023hydraulic}).}
  \label{fig:phenomena}
\end{figure}

In subsurface environments, the injected fluid interacts with the rock matrix, the existing natural fracture (NF) network, and the in-situ stress field (Abe et al., \citeyear{abe2021laboratory}). These interactions dictate the geometry, connectivity, and hydraulic behavior of the resulting fracture network, thereby influencing production efficiency or storage integrity (Warpinski \& Teufel, \citeyear{warpinski1987influence}; Gu \& Weng, \citeyear{gu2012hydraulic}; Zhou et al., \citeyear{zhou2015modeling}). When an advancing hydraulic fracture (HF) encounters a pre-existing NF, several interaction outcomes are possible, including crossing, deflection, arrest, or offset as shown in Figure~\ref{fig:phenomena}(a)-(d). The specific outcome depends on the local stress state, HF-NF orientation, and interface properties. Several criteria have been developed to predict fracture interaction behavior (Renshaw \& Pollard, \citeyear{renshaw1994numerical}; Blanton, \citeyear{blanton1982experimental}; Xu et al., \citeyear{xu2019comprehensive}; Zhao et al., \citeyear{zhao2023hydraulic}). 

However, most of the existing predictive criteria, whether derived from laboratory observations or computational models, are grounded in linear elastic fracture mechanics (LEFM) and are formulated using far-field stress measures. As a result, they inherently assume straight or kinked fracture propagation paths, as illustrated in Figure~\ref{fig:phenomena}(a)-(d). These studies do not explicitly capture local stress perturbations associated with nearby discontinuities or mixed-mode crack-tip loading to reveal physical mechanisms responsible for hydro-mechanical interactions among components of naturally fractured rock masses. For example, the dimensionless stress ratio (Zhao et al.,~\citeyear{zhao2023hydraulic}), shown in Figure~\ref{fig:phenomena}(e), is defined using far-field stresses and does not account for stress redistribution induced by NF slip. In contrast, in real-scale problems, fluid injection induces pore-pressure diffusion and poroelastic stress changes that redistribute local stresses, generate transient shear, and rotate principal stress orientations, ultimately influencing HF trajectories (Berchenko \& Detournay, \citeyear{berchenko1997deviation}; AlTammar et al., \citeyear{altammar2018effect}; Wang et al., \citeyear{wang2022numerical}). Thus, a critical question remains unresolved: \textit{how do local stress perturbations influence HF-NF interaction outcomes?}

In homogeneous, isotropic media, where stresses are uniformly distributed, HFs grow parallel to the maximum principal stress (Hubert \& Willis, \citeyear{hubert1957mechanics}). In-situ stress anisotropy significantly influences fracture geometry, as larger principal stress differences force HFs parallel to $\sigma_{\max}$, whereas near-isotropic conditions promote fractures in multiple orientations, i.e., branched patterns (Zhou et al., \citeyear{zhou2015modeling}). The stress intensity factor near the HF tip decreases with increasing crack length as the local stress state transitions from tensile to compressive (Chuprakov et al., \citeyear{chuprakov2014injection}). The guiding hypothesis of our study is that pre-existing discontinuities near a growing HF locally perturb stresses ahead of the tip, influencing fracture trajectory.

Previous experimental studies on this subject have primarily relied on proxy observations of fracture initiation and propagation, such as pressure history data, tracers, or post-test fracture morphology inspection, rather than simultaneous measurement of the evolving stress–strain state. Advanced imaging techniques have improved detailed deformation characterization, but entail notable limitations. X-ray computed tomography offers three-dimensional visualization of the final internal fracture geometry, yet it cannot capture the transient evolution of stresses and strains. Digital image correlation (DIC) overcomes this constraint by resolving full-field surface displacement and strain localization in near real time (Nguyen et al., \citeyear{nguyen2011fracture}; Gao et al., \citeyear{gao2015investigation}; Pan et al., \citeyear{pan2018digital}; Lattanzi et al., \citeyear{lattanzi2023uncertainty}; Shandilaya \& Roshankhah, \citeyear{shandilaya2024laboratory}). However, due to insufficient spatial resolution at the crack tip, these studies are typically restricted to qualitative crack-tip kinematics and displacement localization, and do not resolve the underlying stress redistribution within the crack process zone that governs fracture trajectory (Sutton, \citeyear{sutton2009image}; Gonzales and Diaz, \citeyear{gomez2025experimental}). To bridge this gap, the present study integrates DIC-derived deformation fields with hydro-mechanically coupled numerical simulations to resolve stress evolution and fracture propagation mechanisms.

Numerical modeling has evolved significantly to address computational challenges of coupled hydro-mechanical fracture propagation in heterogeneous media. Continuum-based approaches employ the finite element method (FEM) to simulate fluid-driven fracture growth. In particular, the extended finite element method (XFEM) enables crack initiation and propagation without predefined crack paths through enrichment of the displacement and pore-pressure fields (Hyman et al., \citeyear{H2016}; Kamali et al., \citeyear{kamali2023modeling}; Sarmadi et al., \citeyear{sarmadi2024phase}). XFEM enriches these fields by introducing additional functions that allow displacement and pressure discontinuities to evolve within standard elements, enabling dynamic fracture nucleation through cohesive traction–separation laws that capture progressive damage (Ortiz \& Pandolfi, \citeyear{ortiz1999finite}; Turon et al., \citeyear{turon2006damage}).

In this study, we investigate the mechanisms governing the interactions between an induced HF and a pre-existing planar, non-cemented, frictional NF under plane-strain conditions by integrating laboratory experiments with numerical simulations. Laboratory experiments employ DIC to measure full-field displacement and strain evolution of the specimens during loading and injection. Corresponding XFEM-based poroelastic simulations explicitly model fracture geometry, fluid flow, and coupled hydraulic fracture propagation to resolve the associated stress-state evolution. 

To this end, the remainder of this paper is organized as follows: Section \ref{sec:experimental} describes the experimental setup, specimen design, boundary conditions, and DIC methodology. Section \ref{sec:numerical} presents the numerical modeling, including coupled hydro-mechanical formulation and XFEM implementation, followed by the calibration and validation in Section \ref{sec:calibration}. Section \ref{sec:results} discusses experimental and numerical results, highlighting mechanistic interpretation of HF-NF interactions across different stress-state regimes. Finally, Section \ref{sec:conclusions} summarizes key findings and their implications for subsurface fracture-network development and reservoir-scale
modeling.

\section{Experimental Modeling}
\label{sec:experimental}
\subsection{Loading and Fluid Injection Test Setup}
\begin{figure}[]
    \centering
    \begin{subfigure}{0.6\textwidth}
        \centering
        \includegraphics[width=\linewidth]{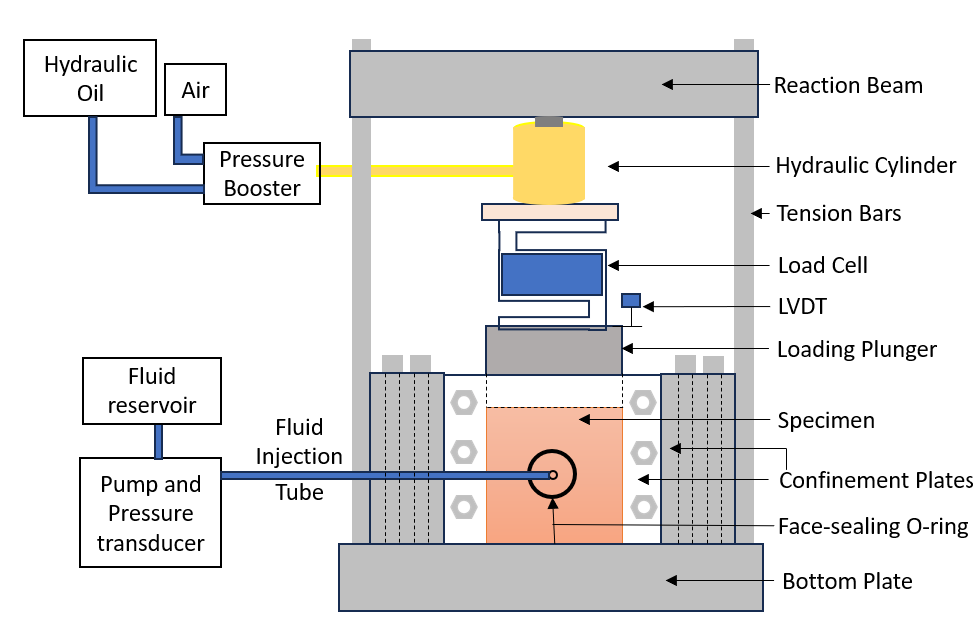}
        \caption{}
        \label{fig:setup}
    \end{subfigure}\hfill
    \begin{subfigure}{0.33\textwidth}
        \centering
        \includegraphics[width=\linewidth]{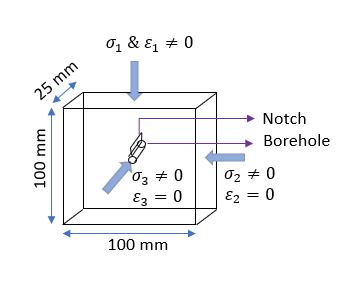}
        \caption{}
        \label{fig:setup-boundary}
    \end{subfigure}
    
    \vspace{1em} 

    \begin{subfigure}{0.31\textwidth}
        \centering
        \includegraphics[width=\linewidth]{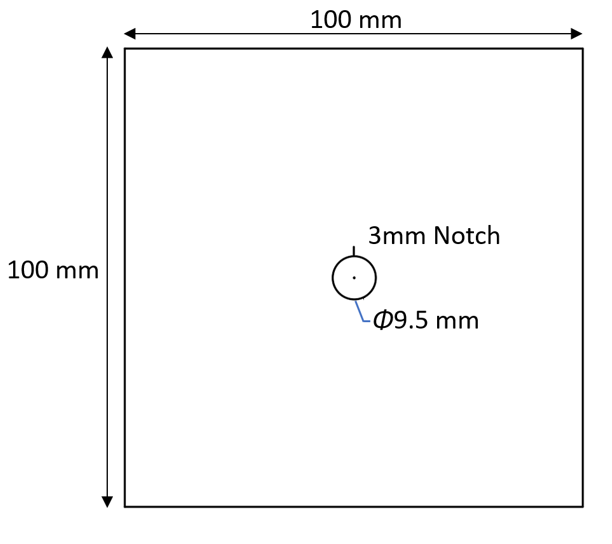}
        \caption{}
        \label{fig:Specimen-3}
    \end{subfigure}
    \begin{subfigure}{0.31\textwidth}
        \centering
        \includegraphics[width=\linewidth]{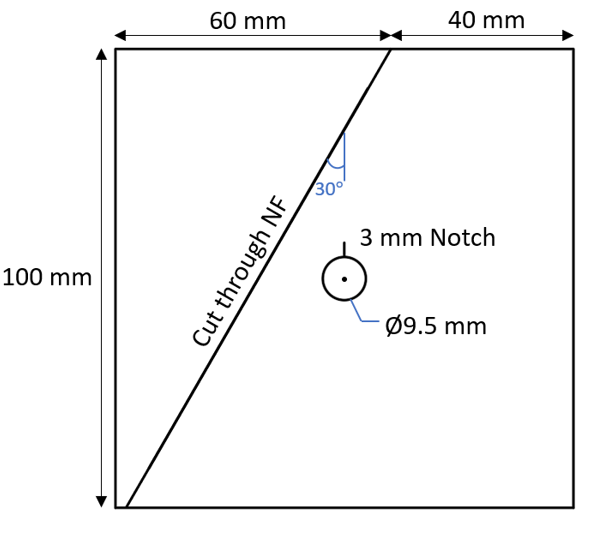}
        \caption{}
        \label{fig:Specimen-2}
    \end{subfigure}
    \begin{subfigure}{0.33\textwidth}
        \centering
        \includegraphics[width=\linewidth]{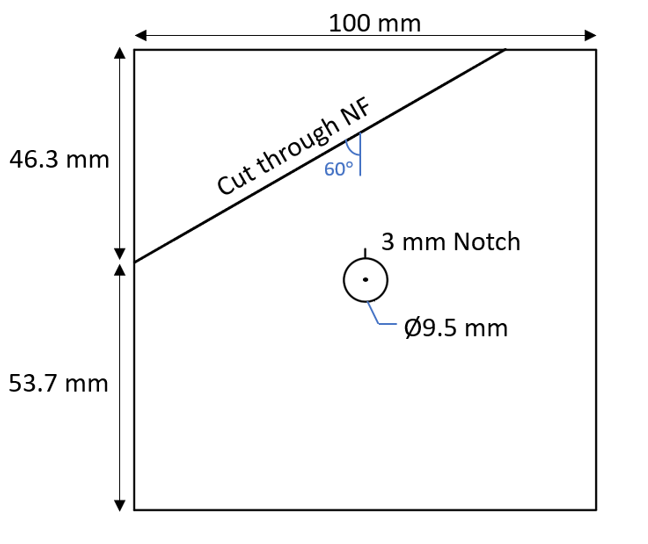}
        \caption{}
        \label{fig:Specimen-1}
    \end{subfigure}
    \caption{Schematic of the experimental setup and test specimens: (a) Loading and injection apparatuses and instrumentation, (b) Mechanical boundary conditions on the test specimen, and 2D cross-sectional geometries of (c) the intact specimen and (d–e) pre-fractured specimens containing a NF inclined at $30^\circ$ and $60^\circ$, respectively, with respect to the maximum principal stress ($\sigma_{\max} = \sigma_{1}$).}
    \label{fig:Exp-setup}
\end{figure}

A two-dimensional schematic view of the loading frame, peripheral equipment, instrumentation, the prismatic test specimens with the machined internal features, and the boundary conditions imposed on the specimen is shown in Figure \ref{fig:Exp-setup}. The loading frame is designed to apply about 12 MPa vertical stress on the test specimen's top side (100 mm $\times$ 25 mm) by a hydraulic cylinder positioned in series between the specimen and the reaction beam. The hydraulic cylinder is fed by a pressure booster. Horizontally, the specimen is securely held between transparent PMMA plates on its large sides (front and back) and aluminum plates on its small sides (left and right) to provide near-zero lateral strain conditions. A high-pressure-rated tube is inserted into the specimen's central borehole and is connected to a reciprocating pump for fluid injection. The setup also includes a load cell to measure the vertical force, an LVDT to record vertical deformations, and a pressure gauge inside the pump to monitor fluid pressure mobilized inside the specimen upon constant flow rate water injection.

\subsection{Materials and Specimen Preparation}
Several prismatic PMMA specimens with overall dimensions of $100 \, \text{mm} \times 100 \, \text{mm} \times 25 \, \text{mm}$ are machined. A borehole with a diameter of 9.5 mm is drilled at the center of the specimens. A 3 mm long blunted-tip notch is fabricated on the top of the borehole wall throughout the thickness using a diamond saw such that the projection of the notch plane is normal to the top side of the specimen. A subset of specimens is then cut in a high-precision machine shop to achieve a single saw cut, i.e., resembling a natural fracture (NF) with its strike 25 mm away from the notch tip. The model NFs are oriented at $30^\circ$ and $60^\circ$ with respect to $\sigma_{\max}$, which is in the vertical direction. 2D cross-sections of the intact and pre-fractured specimens are shown in Figure~\ref{fig:Exp-setup} (c-e)

\subsection{Experimental Procedure}
An experiment on each specimen commences by applying vacuum grease to the regions enclosed by the O-rings on the front and back PMMA confining plates, then positioning the specimen within the confining jacket. Subsequently, all bolts on the confining plates are tightened. The remainder of the setup is assembled in sequence, including the loading plunger, load cell, and hydraulic cylinder. The vertical load is applied through the pressurization of the hydraulic oil in the cylinder through a pressure booster. Water injection into the specimen is executed using a peristaltic Teledyne pump with a constant rate of 9 ml/min, while pressure-time data are recorded. Injection of water into the borehole sealed by the two face-sealing O-rings increases pressure until an HF formed within the PMMA specimens. This is marked by a distinct breaking sound, a sharp drop in the pressure data (see Figure~\ref{fig:pressure-evolution}), and a dip in the force data.

\subsection{Imaging System for Fracture Evaluation}

Two-dimensional DIC is used to monitor the planar deformation and fracture behavior on the specimen surface. The specimen surface is prepared with a random, high-contrast speckle pattern to enable reliable image correlation. Images are acquired using a high-resolution digital camera system and appropriate illumination, with the specimen positioned perpendicular to the camera axis. With a field of view of 76~mm $\times$ 51~mm, a pixel resolution of 9.24 $\micro$m is achieved.

Digital images are acquired at three stages: (i) prior to loading, (ii) after application of the vertical load following a 10-minute equilibration period, and (iii) immediately after the formation of the hydraulic fracture. Surface deformations are computed by correlating consecutive image pairs using a zero-normalized squared differences criterion (Pan et al., \citeyear{pan2009two}). Displacement fields are obtained, and corresponding strain fields are calculated from spatial derivatives of the displacement fields. Images acquired before and after mechanical loading are correlated to identify imposed boundary conditions, while images acquired before and after hydraulic fracture initiation are correlated to quantify deformation and strain evolution during fluid injection. Image correlation is performed using VIC-2D software. Details of the DIC procedure, correlation parameters, and image filtering are provided in Appendix~\ref{sec:DIC}. During the experiments, the vertical force, vertical displacement, and fluid pressure data are also recorded.


\section{Numerical Modeling}
\label{sec:numerical}

\subsection{Governing Equations}
The numerical study employs a fully coupled hydro-mechanical finite element formulation to simulate HF propagation and its interaction with NF. The formulation resolves solid deformation, pore-fluid flow, and fracture evolution within the experimental region of interest.

\begin{figure}[]
\centering 
\includegraphics[width=0.9\textwidth]{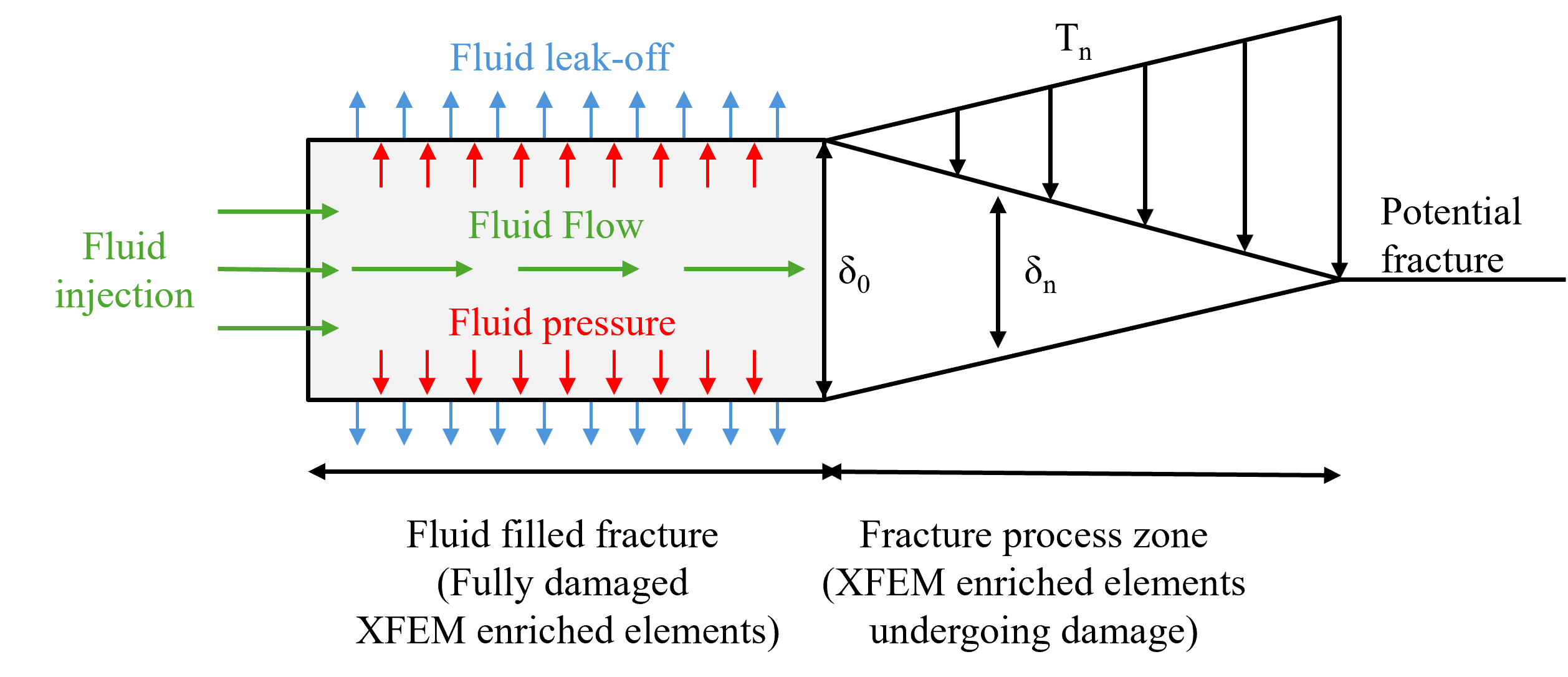} 
\caption{Schematic of the coupled hydro-mechanical processes in hydraulic fracturing, including fluid injection into the fracture, leak-off into the surrounding porous medium, and deformation of the solid matrix.}
\label{fig:fracturing_flow} 
\end{figure}

Hydraulic fracturing in poroelastic media is governed by a nonlinear, coupled system controlling (i) fluid injection and flow within the fracture, (ii) infiltration and pressure diffusion in the surrounding porous matrix, (iii) the mechanical deformation of the host medium, and (iv) fracture initiation and propagation governed by evolving local stress states as illustrated in Figure~\ref{fig:fracturing_flow}. The fluid pressurization drives solid deformation, which in turn alters the fracture aperture and feeds back into flow through the cubic aperture–conductance relation.
(Adachi et al., \citeyear{Ad2007}; Zielonka et al., \citeyear{Zielonka2014}; Hyman et al., \citeyear{H2016}). The governing equations and constitutive relations are summarized below, with full derivations provided in Appendix~\ref{Appendix:Num}.

The hydro-mechanical response of PMMA is modeled using linear poroelasticity under the small deformation assumption. Based on Biot’s theory (\citeyear{biot1941}), the mechanical equilibrium equation is given by (Terzaghi, \citeyear{tr1943}):
\begin{equation}
\nabla \cdot \left( 2G \boldsymbol{\varepsilon} + \lambda\ \text{tr}(\boldsymbol{\varepsilon}) \mathbf{I} - \alpha p \mathbf{I} \right) = \mathbf{0},
\label{eq:poroelastic_equilibrium}
\end{equation}
where $\boldsymbol{\varepsilon}$ is the infinitesimal strain tensor, $p$ is the pore pressure (Pa), $G$ and $\lambda$ are Lamé’s parameters (Pa), $\mathbf{I}$ is the identity tensor, and $\alpha$ is the Biot coefficient. In this poroelastic formulation, it is emphasized that the stress driving fracture propagation is not solely determined by elastic strain. The effective stress includes an explicit contribution from pore pressure. Further details are provided in Appendix~\ref{Appendix:Num}.

Fluid flow in the porous matrix follows Darcy’s law and mass conservation, yielding the pressure diffusion equation (Whitaker, \citeyear{whitaker1986flow}):
\begin{equation}
\frac{1}{M} \frac{\partial p}{\partial t} + \alpha \frac{\partial \varepsilon_v}{\partial t} = 
\nabla \cdot \left( \frac{k}{\mu} \nabla p \right),
\label{eq:matrix_flow}
\end{equation}
where $M$ is the Biot modulus (Pa), $\frac{\partial p}{\partial t}$ is the rate of pressure build-up (Pa/s), $\varepsilon_v = \text{tr}(\boldsymbol{\varepsilon})$ is the volumetric strain, $k$ is the intrinsic permeability (m$^2$), and $\mu$ is the fluid viscosity (Pa$\cdot$s).

Fluid flow within the fracture and its exchange with the surrounding matrix are described using lubrication theory with leak-off, resulting in a nonlinear diffusion equation for the fracture pressure $p_f$ (Pa) (Zielonka et al., \citeyear{Zielonka2014}):
\begin{equation}
\frac{\partial}{\partial s} \left( -\frac{w^3}{12\mu_f} \frac{\partial p_f}{\partial s} \right) 
+ q_{\text{leak}} = Q_{\text{inj}}(s),
\label{eq:fracture_flow}
\end{equation}
where $s$ is the coordinate along the fracture plane (m), $w$ is the local fracture aperture (m), $\mu_f$ is the fracturing-fluid viscosity  (Pa$\cdot$s), $Q_{\text{inj}}$ is the injection rate per unit area (m/s), and $q_{\text{leak}}$ denotes the leak-off flux (m/s).

\begin{figure}[]
\centering
\includegraphics[width=0.5\textwidth]{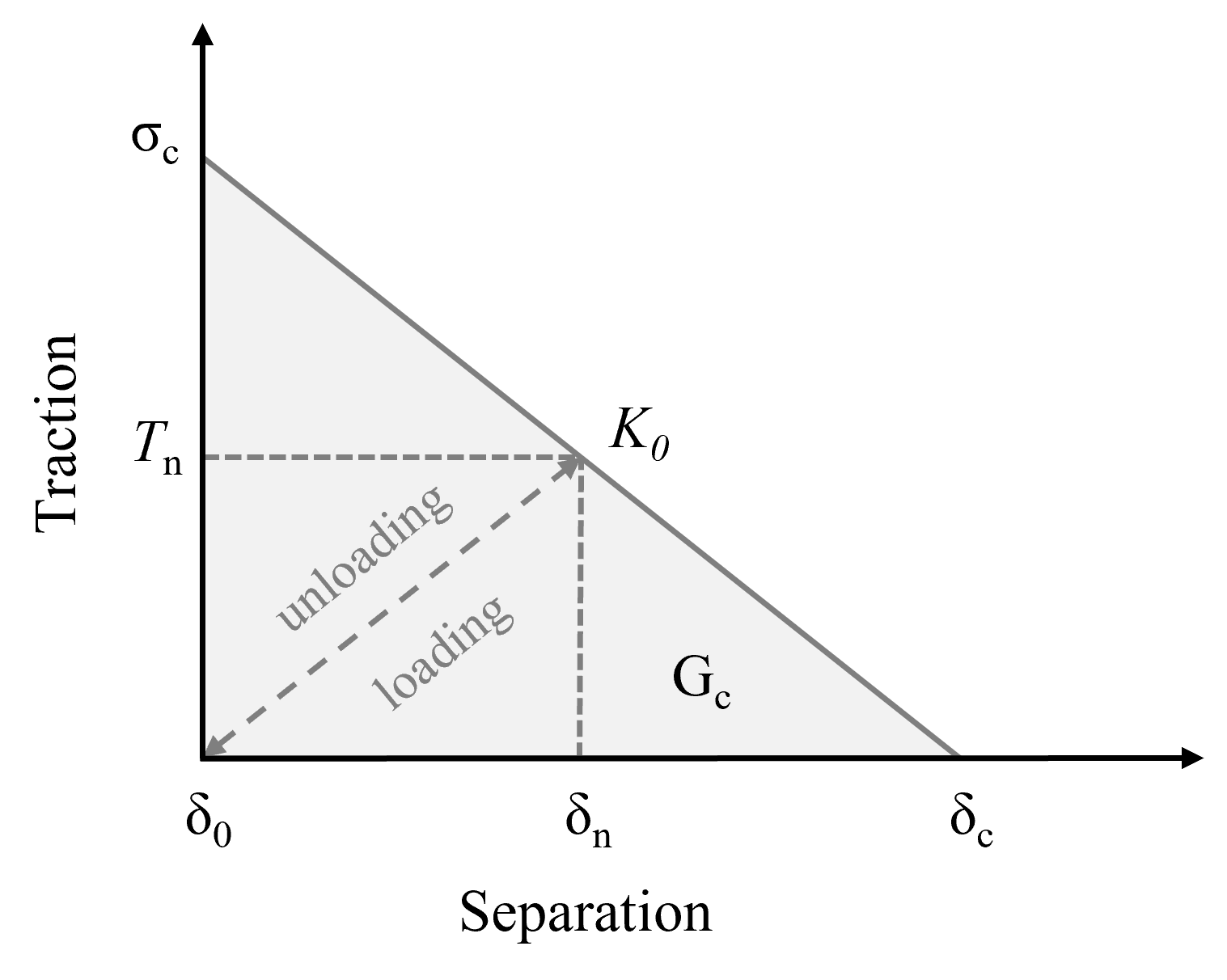}
\caption{Traction–separation law used in the XFEM formulation, showing damage initiation at ($\delta_0$, $\sigma_c$), linear softening to $\delta_c$, and fracture energy $G_c$ as the area under the curve. Unloading path demonstrates permanent displacement discontinuity characteristic of XFEM crack propagation.}
\label{fig:traction}
\end{figure}

The fracture initiation and propagation are modeled using XFEM with cohesive traction-separation law (Barenblatt, \citeyear{Barenblatt1962}; Ortiz \& Pandolfi, \citeyear{Ortiz1999}), as illustrated in Figure~\ref{fig:traction}. The normal traction $T_n$ (Pa) across the fracture varies with the normal separation $\delta_n$ (m) according to:
\begin{equation} 
T_n(\delta_n) = 
\begin{cases} 
K_0 \,\delta_n, & 0 \leq \delta_n \leq \delta_0, \\[4pt] 
\sigma_c \left( 1 - \dfrac{\delta_n - \delta_0}{\delta_c - \delta_0} \right), & \delta_0 < \delta_n \leq \delta_c, \\[6pt]
0, & \delta_n > \delta_c,
\end{cases} 
\label{eq:tsl} 
\end{equation}
where $\sigma_c$ is the cohesive strength (Pa), $\delta_0$ and $\delta_c$ are separations at damage initiation and complete decohesion (m), respectively. The area under the traction-separation curve defines the Mode~I fracture energy $G_{Ic}$ (J/m$^2$). In this study, $\sigma_c$ is set equal to the maximum principal stress. Once initiated, damage evolution follows the Benzeggagh-Kenane (BK) mixed-mode criterion:
\begin{equation}
    G_c = G_{Ic} + \left(G_{IIc} - G_{Ic}\right) 
    \left( \frac{G_{II}}{G_I + G_{II}} \right)^{\beta},
    \label{eq:BK}
\end{equation}
where $G_I$ and $G_{II}$ are the current energy release rates, $G_{Ic}$ and $G_{IIc}$ are the critical values, and $\beta$ is the mode-mixity exponent.

\subsection{Finite Element Models}

All simulations are conducted in \texttt{Abaqus/Standard 2024}, which supports coupled pore-fluid diffusion and stress analysis for hydraulic fracturing. Custom features, such as fluid injection at XFEM-enriched nodes, are implemented through additional \texttt{.inp} keywords following the approach of Zielonka et al. (\citeyear{Zielonka2014}). The ensuing subsections describe the model geometry, boundary conditions, and input parameters used to reproduce the experimental conditions and investigate HF–NF interactions under different local stress states.

\subsubsection{Model Configurations}

A two-dimensional plane-strain model (Figure~\ref{fig:Num_model}) is developed to represent the experimental DIC field of view (76~mm $\times$ 51~mm) of the PMMA specimens, enabling direct spatial comparison between experimental and numerical responses over the same region of interest. The plane-strain assumption is justified because the specimen thickness (25~mm) greatly exceeds the fracture aperture and the applied loading is approximately uniform along the out-of-plane direction. The borehole is modeled as a central circular void of 9.5 mm diameter, with a 3 mm blunted-tip notch introduced on its upper wall to promote fracture initiation.

For pre-fractured configurations, NFs are represented as internal surfaces located 24~mm away from the notch tip and oriented at 30\(^\circ\) or 60\(^\circ\) with respect to the vertical direction (Figure~\ref{fig:Num_model}~(a)). Experimental observations indicated that these saw-cut planes acted primarily as displacement discontinuities, redistributing stresses without noticeable sliding or opening. Since NF interface properties are not available, DIC-derived displacement fields from the loading phase are applied along the NF surfaces as boundary conditions (see Section \ref{sec:BC}). This approach reproduces the experimentally observed deformation and stress perturbations while avoiding the introduction of uncertain interface parameters.

\subsubsection{Mesh and Element Types}

All models are discretized using four-node bilinear pore pressure–displacement elements (CPE4P). Fracture propagation is modeled using XFEM enrichment of these elements, allowing displacement discontinuities across fracture surfaces (i.e., separation and sliding of opposing crack faces without re-meshing) as well as pore-pressure discontinuities (i.e., distinguishing fracture pressure from matrix pore pressure and capturing pressure drops associated with fluid leak-off) to evolve in response to the local stress field.

\begin{figure}[]
\centering 
\includegraphics[width=\textwidth]{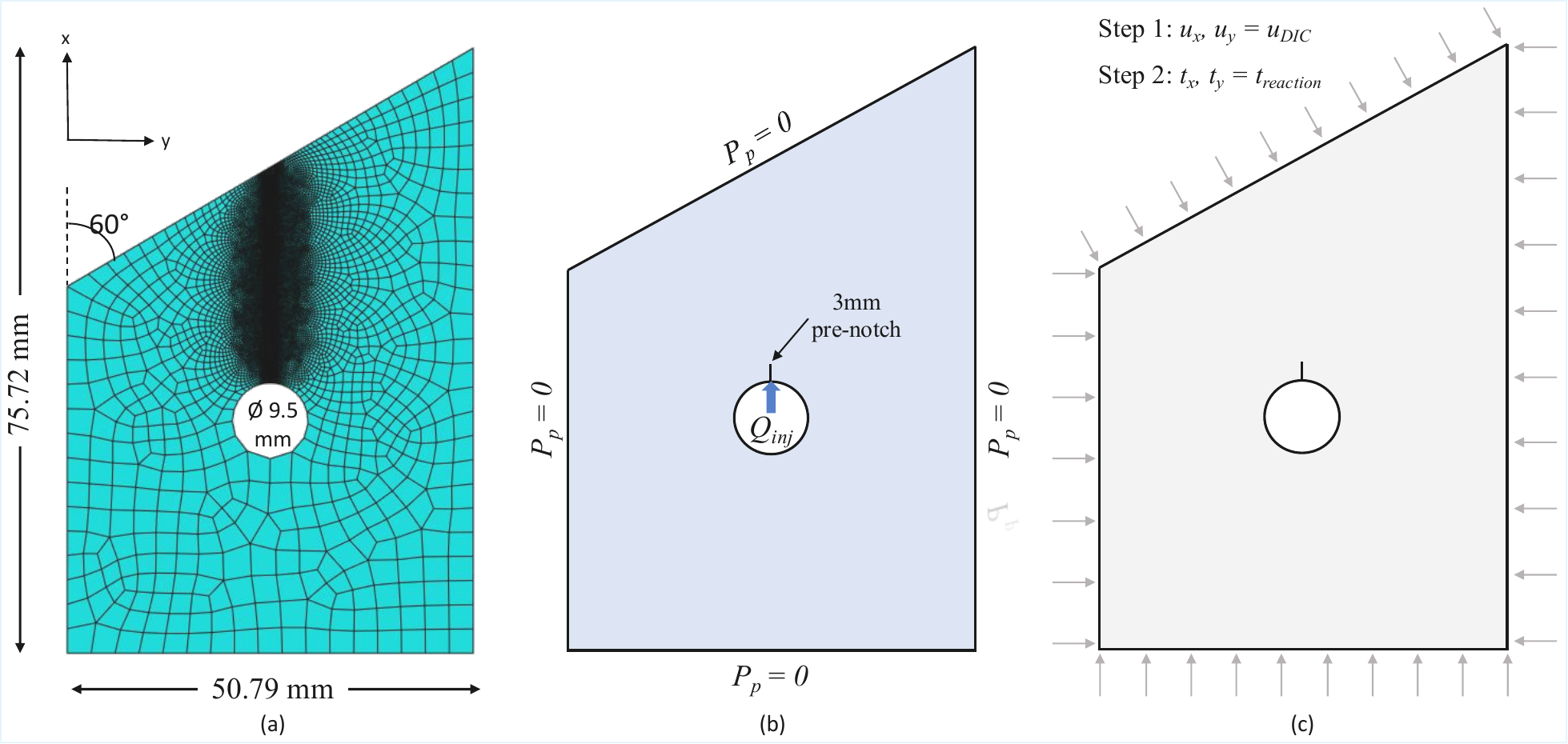}
\caption{Example of finite element model of DIC field of view for PMMA specimen with 60\(^\circ\) Natural Fracture: (a) geometry and mesh discretization with local refinement above the borehole; (b) Hydraulic boundary conditions of zero pore pressure $P_{\text{p}} = 0$ is applied at domain surfaces and $Q_{inj}$ is the fluid injected at the notch above the borehole; (c) Mechanical boundary conditions include Step 1, where $u_x \text{ and }~ u_y$ are the displacements derived from DIC analysis applied at domain nodes and Step 2, where $t_x \text{ and }~ t_y$ are the equivalent reaction tractions.} 
\label{fig:Num_model} 
\end{figure}

Local mesh refinement is applied near the pre-notch and along the NF boundaries to resolve steep stress gradients. The refinement follows the criterion proposed by Turon et al. (\citeyear{turon2006damage}):
\begin{equation}
h_e < \frac{9E \, G_c}{8\sigma_c^2},
\end{equation}
which yields a target element size of approximately 0.275~mm. To ensure numerical stability and adequate resolution of the fracture tip, elements with a characteristic size of 0.1~mm are used near the notch and along the expected crack trajectory above the borehole, as shown in Figure~\ref{fig:Num_model}(a).

\subsubsection{Initial and Boundary Conditions}
\label{sec:BC}
The model is initialized with zero pore pressure, in-situ stress, and saturation to match the experimental starting conditions. External boundaries are maintained at zero pore pressure throughout the analysis to represent drainage and ambient laboratory pressure conditions, as demonstrated in Figure~\ref{fig:Num_model}(b).

To impose experimentally consistent mechanical loading, a two-step boundary condition strategy based on DIC measurements is adopted as shown in Figure~\ref{fig:Num_model}(c). Experimental displacement fields obtained from DIC are transformed from image coordinates to physical coordinates and are mapped into the Abaqus global coordinate system, which defines the fixed Cartesian coordinate system of the finite element mesh nodal locations. The displacement fields are then interpolated onto finite element nodal locations using nearest-neighbor interpolation and are applied as boundary conditions despite differences in spatial resolution.

In Step~1, the interpolated DIC displacement components ($u_x$, $u_y$) are prescribed along the outer boundaries to reproduce the experimental deformation field and to compute the corresponding reaction forces. In Step~2, these reaction forces are reapplied as equivalent boundary tractions, allowing the specimen to deform freely during fluid injection while maintaining the experimentally induced stress state. Hydraulic fracturing is initiated by prescribing a constant flow rate of $1.5\times 10^{-7}$~m$^3$/s (equivalent to the experimental flow rate of 9~mL/min) at the pre-notch above the borehole (Figure~\ref{fig:Num_model}(b)).

\subsubsection{Model Parameters}

PMMA is modeled as an isotropic, linear poroelastic material with parameters obtained from the literature and adjusted to match the experimental displacement and strain responses after mechanical loading. Parameters without direct literature values, including effective porosity and leak-off coefficients, are calibrated to reproduce the experimentally observed pressure–time response following fracture initiation. These model parameters are summarized in Table~\ref{tab:material-properties} with their calibration and validation described in the following section.

\begin{table}[]
\centering
\caption{Model parameters used in the numerical simulations.}
\label{tab:material-properties}
\begin{tabularx}{0.9\textwidth}{l c >{\RaggedRight\arraybackslash}X}
\toprule
\textbf{Parameter} & \textbf{Value} & \textbf{References} \\
\midrule
\multicolumn{3}{l}{\textit{Solid Matrix (PMMA)}} \\
Young’s modulus, $E$ (MPa) & 3300 &  Zhang et al., \citeyear{Zhang2015} \\
Poisson’s ratio, $\nu$  & 0.37 & Khadraoui et al., \citeyear{Khadraoui2021}  \\
Density, $\rho$ (kg/m$^3$) & 1,180 &  Zhang et al., \citeyear{Zhang2015}; Khadraoui et al., \citeyear{Khadraoui2021} \\
Biot coefficient, $\alpha$ & $\sim$1.0 & Carrillo et al. \citeyear{carrillo2019darcy} \\
Porosity, $\phi$ & 0.01 &  Calibrated in this study \\ \\
Hydraulic conductivity, $K$ (m/s) & $1.6 \times 10^{-12}$ & Shimko et al., \citeyear{shimko2007development}\\
Mode~I fracture energy, $G_{Ic}$ (J/m$^2$) & 350 &  Zhang et al., \citeyear{Zhang2015} \\
Mode~II fracture energy, $G_{IIc}$ (J/m$^2$) & 700 & Calibrated in this study \\  \\
BK exponent, $\beta$ & 1.5 & Calibrated in this study \\
Maximum principal stress, $\sigma_{c}$ (MPa) & 60 &  Zhang et al., \citeyear{Zhang2015} \\
\midrule
\multicolumn{3}{l}{\textit{Fracturing Fluid (Water)}} \\
Viscosity, $\mu$ (Pa$\cdot$s) & 0.001 & Zielonka et al., \citeyear{Zielonka2014} \\
Specific weight, $\gamma$ (N/m$^3$) & 9800 & Zielonka et al., \citeyear{Zielonka2014}  \\
\midrule
\multicolumn{3}{l}{\textit{Leak-off Coefficients}} \\
Top fracture face, $C_{t}$ (m/s/Pa) & $6 \times 10^{-11}$ & Calibrated in this study \\
Bottom fracture face, $C_{b}$ (m/s/Pa) & $6 \times 10^{-11}$ & Calibrated in this study \\
\bottomrule
\end{tabularx}
\end{table}

\section{Model Calibration and Validation}
\label{sec:calibration}

\subsection{Calibration Strategy and Parameter Selection}
Interpreting HF–NF interaction mechanisms requires a numerical framework that reproduces experimentally observed deformation and pressure responses while resolving internal stress fields that cannot be measured directly by DIC. In this study, an XFEM-based poroelastic model is calibrated using a subset of the experimental data and subsequently assessed for its predictive capability across all configurations.

Model calibration follows the physical sequence of the experiments. Mechanical calibration is first performed using DIC-measured displacement and strain fields obtained after mechanical loading for the intact and 60$^\circ$ NF configurations. Correspondingly, these two cases represent a purely tensile stress regime and a mixed-mode stress state induced by an inclined NF. Matching the post-loading deformation fields ensures that the elastic properties and imposed boundary conditions reproduce the correct pre-fracture stress state, which governs subsequent fracture initiation and propagation. Hydraulic calibration is then carried out using the experimentally measured pressure–time responses of the same two specimens during HF. The breakdown pressure and post-breakdown pressure evolution are used to constrain parameters governing fluid–solid interaction, including fracture energy and leak-off behavior. 

Small, but finite, porosity and leak-off are introduced to enable coupled poroelastic behavior and to reproduce the experimentally observed post-breakdown pressure decay. These parameters do not imply measurable seepage in the experiments, but rather provide a physically consistent regularization of the coupled hydro-mechanical problem. The selected porosity and leak-off coefficients lie within ranges commonly adopted for thermoplastic polymers and other low-permeability materials in HF simulations (Cruz et al., \citeyear{CRUZ2018385}). 

Fracture energy and BK exponent parameters are guided by reported values for PMMA and similar amorphous thermoplastic polymers. Mode~I fracture energies for PMMA are typically reported in the range of 300–500~J/m$^2$ (Zhang et al., \citeyear{Zhang2015}), while Mode~II fracture energies are commonly taken as one to two times higher to reflect increased resistance under shear-dominated fracture (Sun et al., \citeyear{SUN2024e37438}). Additionally, mode-mixity exponents generally fall within the range $\beta=1.0-2.0$, indicating a nonlinear interaction between Mode~I and Mode~II fracture energies under mixed-mode loading (Blackman, \citeyear{BLACKMAN2023579}). All the values adopted in this study fall within these reported bounds and are further constrained through calibration against the measured pressure–time responses and fracture trajectory. 

\subsection{Validation and Predictive Capability}

Model validation assesses whether the calibrated model reproduces experimental observations that are not directly targeted during parameter fitting, including spatial deformation patterns, fracture trajectories, and stress redistribution characteristics. Validation is performed using the intact and 60$^\circ$ NF specimens, while the 30$^\circ$ NF is reserved for fully predictive assessment.

\begin{figure}[]
  \centering
  \includegraphics[width=0.8\textwidth]{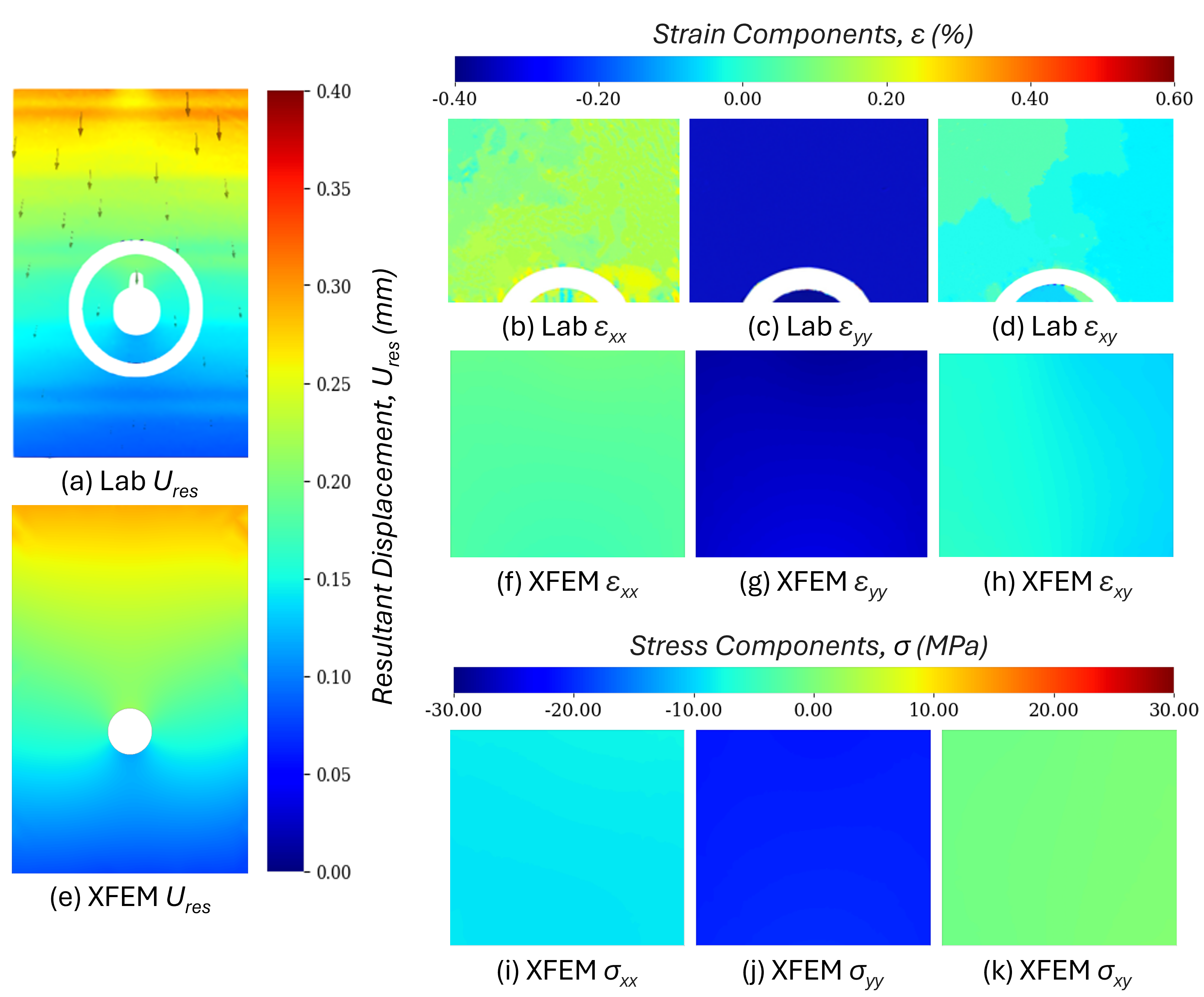}
  \caption{Displacement, strain, and stress fields for the intact PMMA specimen under plane-strain mechanical loading. Experimental results obtained from DIC are shown in panels (a–d), and corresponding numerical results from XFEM are shown in panels (e–k). The top row shows DIC results: (a) resultant displacement in the 76~mm × 51~mm field of view and (b-d) horizontal, vertical, and shear strain fields in the region above the notch. Note: Borehole, notch, and O-ring regions are excluded from the DIC analyses. Middle and bottom rows are XFEM results: (e) resultant displacement in the 76~mm × 51~mm field of view, (f‒h) horizontal, vertical, and shear strains, and (i‒k) horizontal, vertical, and shear stresses in the region above the borehole.}
  \label{fig:baseline-Mechanical}
\end{figure}

After mechanical loading, the predicted displacement and strain fields closely match DIC measurements for both calibrated configurations. For the intact specimen, Figure~\ref{fig:baseline-Mechanical} shows that the model reproduces the magnitude and spatial distribution of deformation ($u$) within approximately 10\% of the experimental measurements. In particular, it captures the localized tensile horizontal strain above the notch ($\varepsilon_{xx} \approx 0.18\%$), the uniform vertical strain compression ($\varepsilon_{yy} \approx -0.35\%$), and the absence of significant shear strain. These results confirm that the pre-fracture stress state remains aligned with the far-field maximum principal stress, $\sigma_{\max}$, and is dominated by pure tensile opening.

\begin{figure}[]
  \centering
  \includegraphics[width=0.8\textwidth]{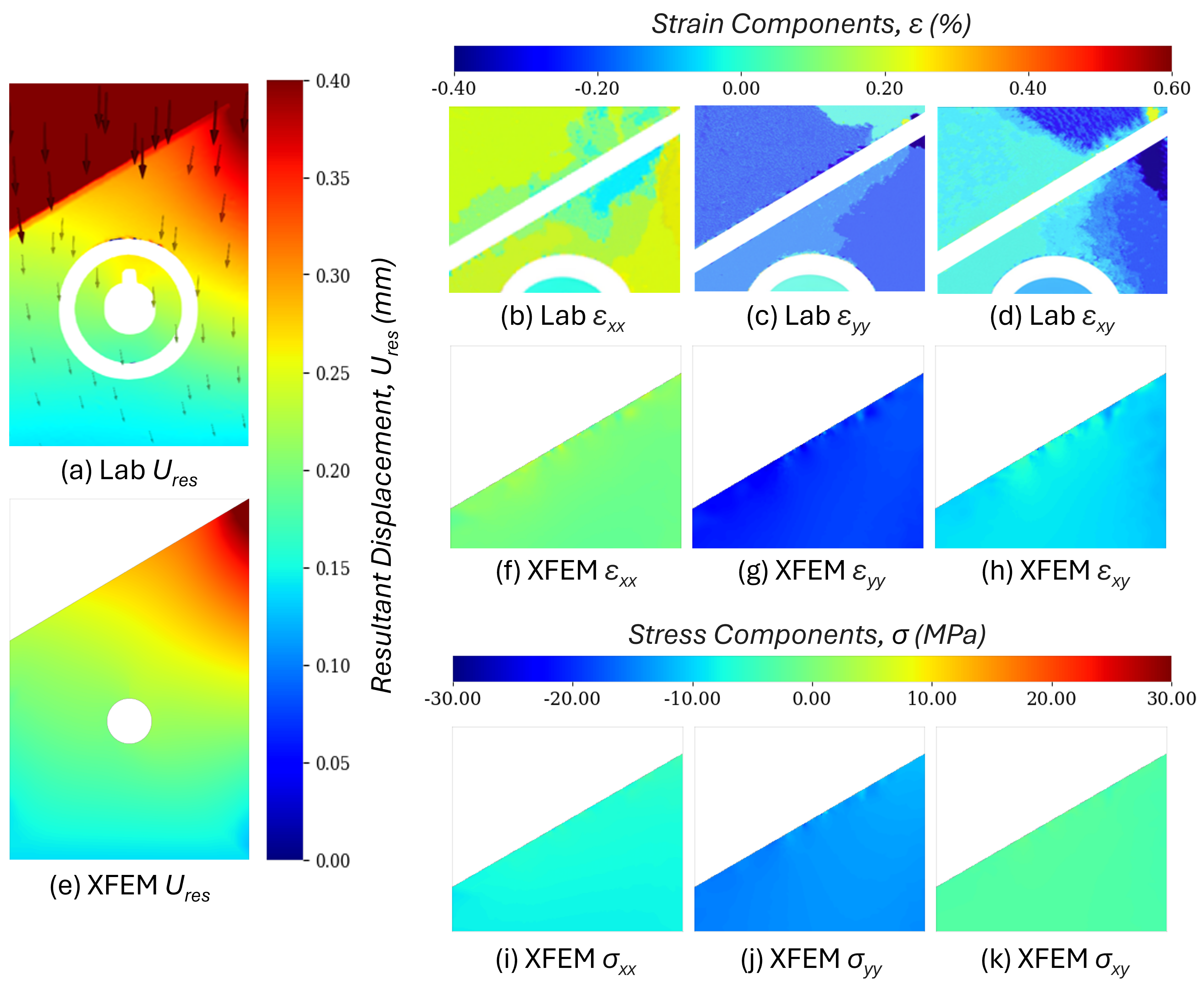}
  \caption{Displacement, strain, and stress fields for the 60$^\circ$ NF configuration PMMA specimen under plane-strain mechanical loading. Experimental results obtained from DIC are shown in panels (a–d), and corresponding numerical results from XFEM are shown in panels (e–k). Top row are DIC results: (a) resultant displacement in the 76~mm × 51~mm field of view and (b-d) horizontal, vertical, and shear strain fields in the region above the notch. Note: Borehole, notch, and O-ring regions are excluded from the DIC analyses. Middle and bottom rows are XFEM results: (e) resultant displacement  in the 76~mm × 51~mm field of view, (f‒h) horizontal, vertical, and shear strains, and (i‒k) horizontal, vertical, and shear stresses in the region above the borehole.}
  \label{fig:mixed-mechanical-60}
\end{figure}

For the 60$^\circ$ NF configuration, Figure~\ref{fig:mixed-mechanical-60} reveals that the model reproduces the asymmetric displacement and strain patterns induced by the inclined NF. In this case, localized shear strain develops with a consistent sign relative to the far-field $\sigma_{\max}$, indicating counter-clockwise rotation of the local stress field toward the NF. The correct sign and spatial localization of shear strain ($\varepsilon_{xy}$) demonstrate that the model captures NF-induced stress redistribution during mechanical loading, which is critical for subsequent HF behavior.


\begin{figure}[]
  \centering
  \includegraphics[width=1\textwidth]{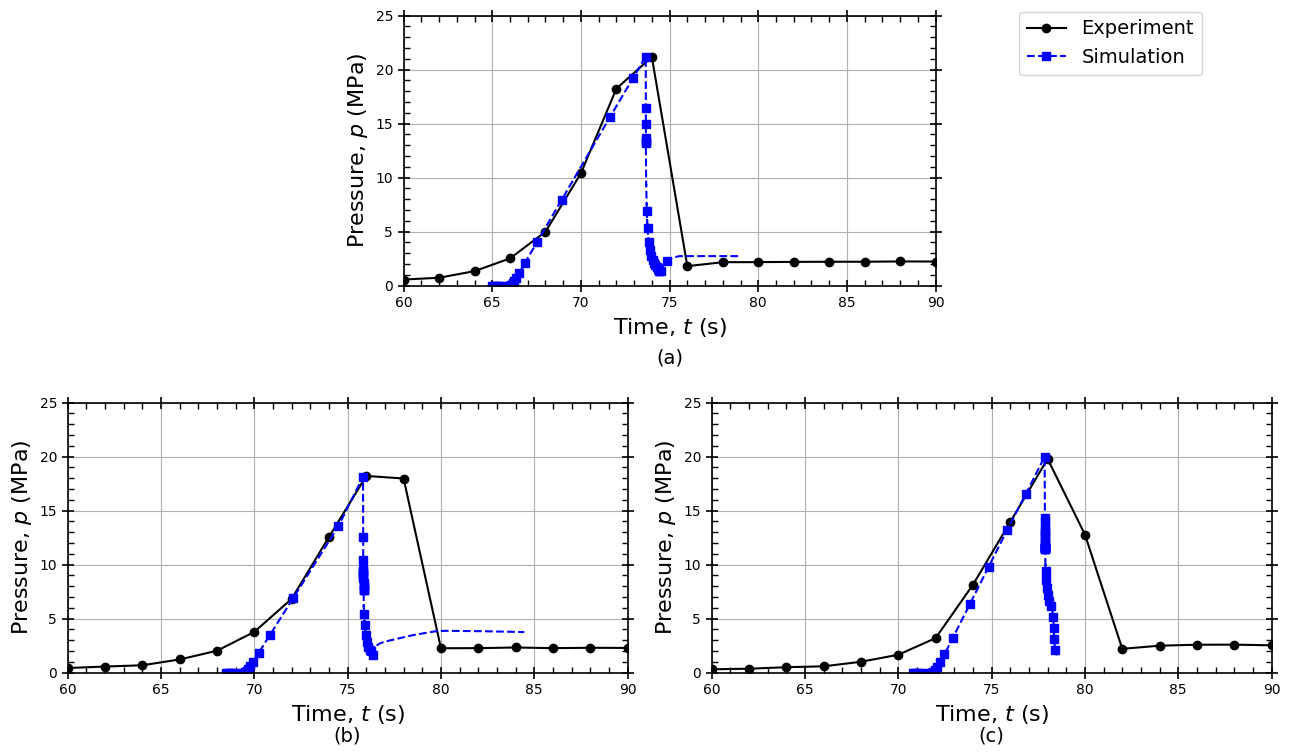}
  \caption{Experimental and numerical pore pressure evolution during hydraulic fracturing. (a) Intact specimen, (b) 60$^\circ$ NF configuration, (c) 30$^\circ$ NF configuration.}
  \label{fig:pressure-evolution}
\end{figure}

\newpage
With the mechanically calibrated stress state established, the model reproduces the experimental pressure–time response during hydraulic fracturing. For the intact specimen, the simulated breakdown pressure of 21.3~MPa and post-breakdown pressure plateau of approximately 2.5~MPa closely match the experimental measurements (Figure~\ref{fig:pressure-evolution}(a)). For the 60$^\circ$ NF configuration, the model captures the reduced breakdown pressure and modified pressure evolution resulting from NF-induced relaxation of the normal stress acting at the notch (Figure~\ref{fig:pressure-evolution}(b)). In both cases, the simulations follow the observed fracture trajectories, including straight vertical propagation aligned with $\sigma_{\max}$ in the intact specimen and deflected propagation toward the NF in the 60$^\circ$ inclined configuration.

Minor discrepancies between experimental and numerical results are confined to regions near the borehole and NF interfaces, where steep displacement gradients occur. These differences likely arise from (i) viscoelastic relaxation of PMMA during equilibration, (ii) drilling-induced surface roughness and microcracking along saw-cut NFs, and (iii) uncertainty associated with extrapolating DIC-derived displacement fields onto the finite element mesh near the NFs. Additionally, the NF is represented geometrically without an explicit interface constitutive law, which limits accuracy in its immediate vicinity. Despite these localized effects, the model consistently reproduces the measured deformation fields, pressure evolution, as well as fracture paths.

Following calibration and validation, the numerical model is applied without further parameter adjustment to predict the hydro-mechanical response of all configurations. In particular, the 30$^\circ$ NF configuration is treated as a fully predictive case. In this setting, the model resolves internal stress redistribution, shear development relative to the far-field stress orientation, and mixed-mode fracture propagation mechanisms that are not directly accessible from the experiments.



\newpage
\section{Results and Discussion}
\label{sec:results}
This section presents and discusses the experimental and numerical results of hydraulic fracturing tests performed on intact and pre-fractured PMMA specimens under plane-strain conditions. Full-field DIC measurements are used to characterize surface deformation and strain evolution, while corresponding XFEM simulations resolve the internal stress fields governing fracture initiation, deflection, and mode transition. Results are first presented for the intact specimen to establish a reference Mode~I fracture regime, followed by pre-fractured configurations that illustrate how inclined NFs modify local stress states and drive mixed-mode HF propagation.

\subsection{Reference Stress Regime: Mode I Propagation}
\label{Sec:tensile-intact}

\subsubsection{Baseline Mechanical Loading and Notch-Induced Stress Concentration}
The intact specimen is analyzed to isolate the mechanical response and stress distribution in the absence of pre-existing natural fractures. Under the imposed plane-strain boundary conditions, the experimentally measured displacement field obtained from DIC is shown in Figure~\ref{fig:baseline-Mechanical}(a), with the corresponding XFEM prediction shown in Figure~\ref{fig:baseline-Mechanical}(e). The observed displacement patterns are consistent with the expected plane-strain response of a laterally confined specimen, characterized by decreasing vertical displacement toward the fixed base due to side friction and small, symmetric horizontal compression induced by lateral confinement. The numerical results closely reproduce these trends, confirming that the applied boundary conditions accurately represent the experimental mechanical state.

Figures~\ref{fig:baseline-Mechanical}(b-d) present the DIC strain fields in the region right above the borehole. The $\varepsilon_{xx}$ field (Figure~\ref{fig:baseline-Mechanical}(b)) reveals a localized tensile zone above the notch of approximately 0.18\%, arising from stress concentration associated with the borehole geometry, while the surrounding region remains largely undeformed. The $\varepsilon_{yy}$ field (Figure~\ref{fig:baseline-Mechanical}(c)) exhibits uniform compression of approximately $-0.35\%$ across the specimen, confirming dominant vertical loading. The $\varepsilon_{xy}$ field (Figure~\ref{fig:baseline-Mechanical}(d)) remains negligible throughout the domain. This combination of localized tensile $\varepsilon_{xx}$ above the notch, globally compressive field $\varepsilon_{yy}$, and near-zero $\varepsilon_{xy}$, creates a purely Mode~I fracture initiation condition that favors vertical crack propagation.

For the same region, the XFEM simulation reproduces the experimentally observed displacement and strain fields, as evident in Figures~\ref{fig:baseline-Mechanical}(f–h). The stress fields in Figures~\ref{fig:baseline-Mechanical}(i–k) show spatially uniform far-field stresses, with dominant vertical compressive stress ($\sigma_{yy}$, Figure~\ref{fig:baseline-Mechanical}(j)), horizontal compressive stress ($\sigma_{xx}$, Figure~\ref{fig:baseline-Mechanical}(i)) approximately one-third of $\sigma_{yy}$, and negligible shear stress ($\sigma_{xy}$, Figure~\ref{fig:baseline-Mechanical}(k)).

Overall, the intact specimen establishes an unperturbed stress–strain and pressure–time response governed by tensile opening aligned with the far-field $\sigma_{\max}$. This reference state provides a baseline for interpreting NF-induced stress redistribution and mixed-mode fracture behavior in the pre-fractured configurations that follow.
\subsubsection{Hydraulic Fracture Initiation and Propagation}

\begin{figure}[]
  \centering
  \includegraphics[width=0.98\textwidth]{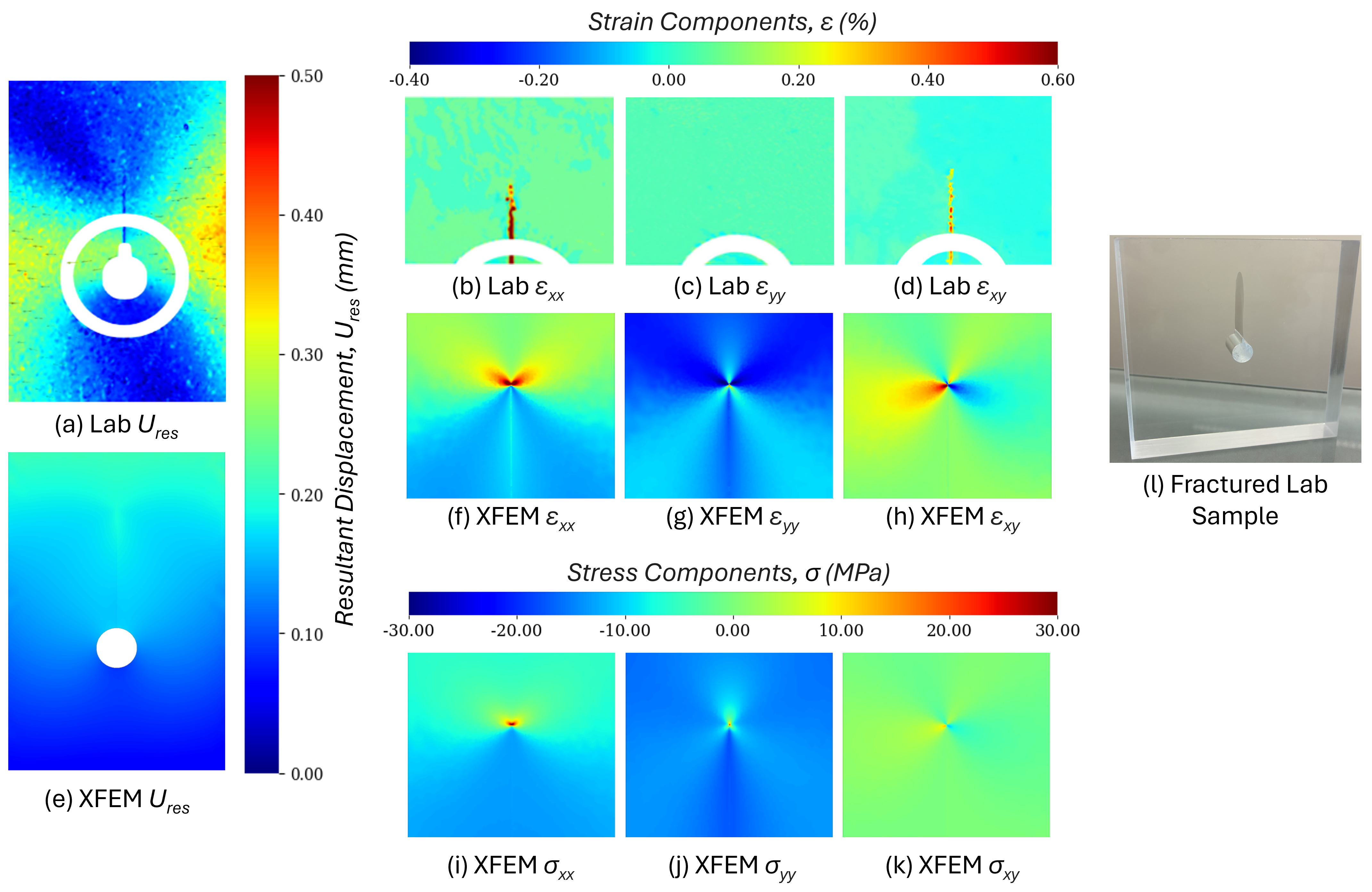}
  \caption{Displacement, strain, and stress fields for the intact PMMA specimen after hydraulic fracture. Experimental results obtained from DIC are shown in panels (a–d), and corresponding numerical results from XFEM are shown in panels (e–k). The top row shows DIC results: (a) resultant displacement in the 76~mm × 51~mm  field of view and (b-d) horizontal, vertical, and shear strain fields in the region above the notch. Note: Borehole, notch, and O-ring regions are excluded from the DIC analyses. Middle and bottom rows are XFEM results: (e) resultant displacement in the 76~mm × 51~mm field of view, (f‒h) horizontal, vertical, and shear strains, and (i‒k) horizontal, vertical, and shear stresses in the region above the borehole. (l) A digital image of a hydraulically fractured intact PMMA specimen.}
  \label{fig:baseline-HF}
\end{figure}

After mechanical loading, water injection at a constant rate of 9~mL/min induced an HF at a pressure of 21.3~MPa in the intact PMMA specimen, followed by a rapid pressure drop to approximately 2.5~MPa as shown in Figure~\ref{fig:pressure-evolution}(a). The abrupt pressure drop and subsequent stable plateau indicate sustained HF propagation with minimal fluid leak-off, consistent with the low-permeability poroelastic response represented in the numerical model. Post-fracture imaging, DIC measurements, and corresponding XFEM results (Figure~\ref{fig:baseline-HF}(a–l)), all show a single, continuous vertical HF extending from the notch toward the top boundary through the specimen thickness, with no deviation or branching. This straight vertical trajectory confirms that HF propagation occurs perpendicular to the minimum principal stress and parallel to the maximum principal stress, consistent with Mode~I fracture mechanics (Anderson et al., \citeyear{anderson2017fracture}).

The strain fields show dominant tensile localization in $\varepsilon_{xx}$ along the fracture path, reaching values of approximately $0.45\%$ (Figure~\ref{fig:baseline-HF}(b,f)). In the DIC measurements, both $\varepsilon_{yy}$ and $\varepsilon_{xy}$ remain close to zero across most of the field of view (Figure~\ref{fig:baseline-HF}(c,d)), indicating negligible vertical and shear deformation at the experimental scale. The XFEM results resolve small, highly localized $\varepsilon_{yy}$ and $\varepsilon_{xy}$ components near the crack tip during propagation. In Figure~\ref{fig:baseline-HF}(g), $\varepsilon_{yy}$ exhibits localized tensile concentrations at the crack tip due to fluid-driven crack opening, while remaining compressive in the far field as a result of the imposed mechanical loading. In contrast, in Figure~\ref{fig:baseline-HF}(h), the $\varepsilon_{xy}$ shows antisymmetric lobes with opposite signs on either side of the fracture, resulting in no net shear contribution to crack propagation and preserving a Mode~I opening response.

The XFEM model successfully reproduced the experimental response, capturing both HF initiation and propagation behavior, visible in Figures~\ref{fig:baseline-HF}(e)-(k). As predicted by the LEFM criterion, which incorporates fracture toughness, Poisson’s ratio, and the applied far-field stress state, the model yields a breakdown pressure of 21.3~MPa, in close agreement with the experiments (Roshankhah, et al. \citeyear{roshankhah2018neutron}). The simulated HF follows the same vertical trajectory observed experimentally. The computed deformation fields also mirror the DIC measurements, with localized tensile $\varepsilon_{xx}$ (Figure~\ref{fig:baseline-HF}(b) and (f)) along the fracture path and negligible shear strain $\varepsilon_{xy}$ (Figure~\ref{fig:baseline-HF}(d) and (h)).

The associated stress fields (Figure~\ref{fig:baseline-HF}(i–k)) show tensile $\sigma_{xx}$ concentrated near the fracture surfaces, compressive $\sigma_{yy}$ in the far field, and $\sigma_{xy}$ confined to the immediate crack-tip region. This stress distribution is characteristic of Mode~I fracture propagation, where fracture opening is governed by tensile normal stresses with negligible shear.


\subsection{Mixed-Mode Fracture Behavior in Pre-fractured Specimens}
\subsubsection{Stress Redistribution in Pre-fractured Specimens During Mechanical Loading}

\begin{figure}[]
  \centering
  \includegraphics[width=0.8\textwidth]{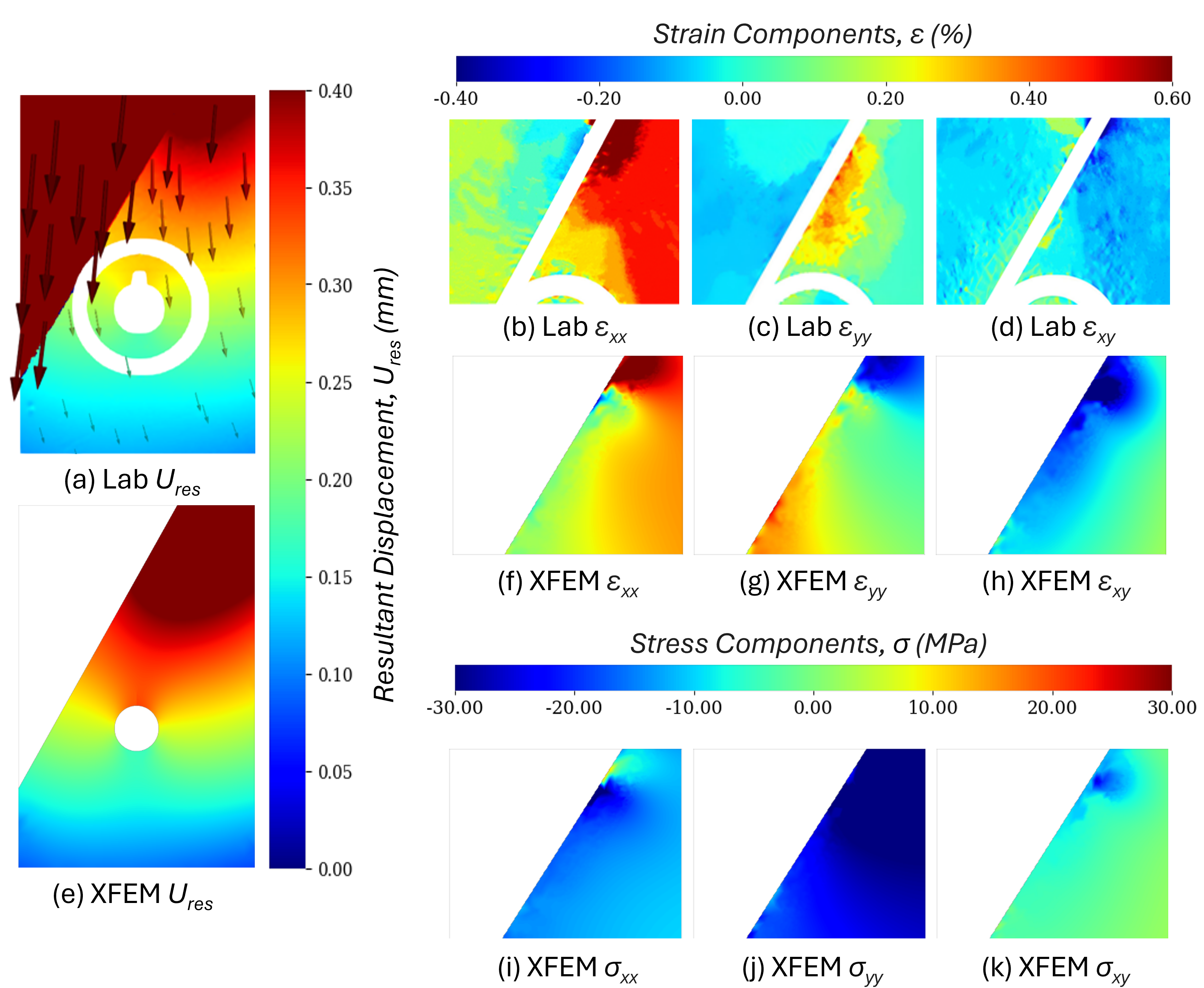}
  \caption{Displacement, strain, and stress fields for the 30$^\circ$ NF configuration PMMA specimen under plane-strain mechanical loading. Experimental results obtained from DIC are shown in panels (a–d), and corresponding numerical results from XFEM are shown in panels (e–k). Top row are DIC results: (a) resultant displacement in the 76~mm × 51~mm field of view and (b-d) horizontal, vertical, and shear strain fields in the region above the notch. Note: Borehole, notch, and O-ring regions are excluded from the DIC analyses. Middle and bottom rows are XFEM results: (e) resultant displacement  in the 76~mm × 51~mm field of view, (f‒h) horizontal, vertical, and shear strains, and (i‒k) horizontal, vertical, and shear stresses in the region above the borehole.}
  \label{fig:mixed-mechanical-30}
\end{figure}

The full-field surface displacement and strain responses for the specimen containing inclined NFs are presented in Figures~\ref{fig:mixed-mechanical-60} and \ref{fig:mixed-mechanical-30}. In contrast to the intact specimen, both pre-fractured configurations exhibit strongly asymmetric deformation during mechanical loading, due to the presence of the NF interface. For the 30$^\circ$ NF configuration, Figures~\ref{fig:mixed-mechanical-30}(a) and (e) show pronounced compressive deformation localized below the NF and on the left side of the notch. On the other hand, for the 60$^\circ$ configuration, Figures~\ref{fig:mixed-mechanical-60}(a) and (e) indicate that compressive deformation is concentrated on the right side of the notch below the NF. These asymmetric displacement patterns demonstrate that inclined NFs perturb the far-field stress state even prior to fluid injection, reflecting stress redistribution associated with NF readjustment under vertical compression when compared to the baseline specimen.

The DIC strain fields further reveal a distinct difference in the magnitude, sign, and distribution of $\varepsilon_{xy}$ between the two NF configurations during mechanical loading. In the 30$^\circ$ NF specimen, Figures~\ref{fig:mixed-mechanical-30}(b–d) show localized positive $\varepsilon_{xy}$ of approximately 0.10\% developing above the notch and below the NF plane. This $\varepsilon_{xy}$ localization is accompanied by tensile $\varepsilon_{xx}$ of $\sim0.3\%$ on the right side of the NF above the notch (Figure~\ref{fig:mixed-mechanical-30}(b)) and a tensile $\varepsilon_{yy}$ band of $\sim0.2\%$ around concentrated along the NF plane (Figure~\ref{fig:mixed-mechanical-30}(c)). Together, these strain components indicate asymmetric deformation associated with shear-driven rotation of local stresses away from the NF. 

On the contrary, the 60$^\circ$ NF specimen exhibits an opposite shear response. DIC strain fields in Figures~\ref{fig:mixed-mechanical-60}(b–d) show localized negative $\varepsilon_{xy}$ of about $-0.07\%$ above the notch, indicating a contrasting sense of shear relative to the 30$^\circ$ configuration. The corresponding $\varepsilon_{xx}$ field (Figure~\ref{fig:mixed-mechanical-60}(b)) exhibits slight compression of approximately $-0.05\%$ near the notch, transitioning to mild tensile strain of approximately 0.2\% farther from the notch below the NF. Unlike the 30$^\circ$ specimen, the $\varepsilon_{yy}$ field the 60$^\circ$ configuration (Figure~\ref{fig:mixed-mechanical-60}(c)) remains compressive. Overall, these differences in strain patterns between the pre-fractured configurations govern the local stress states prior to fluid injection and subsequent HF propagation.

The XFEM simulations reproduce the asymmetric deformation and strain patterns observed experimentally (Figures~\ref{fig:mixed-mechanical-30}(e-h) and \ref{fig:mixed-mechanical-60}(e-h)), capturing displacement magnitudes within approximately 18\% and strain distributions within 20\% for both configurations. The computed stress fields further confirm the strain-based observations, with a slightly positive shear stress component $\sigma_{xy}$ in the 30$^\circ$ NF configuration (Figures~\ref{fig:mixed-mechanical-30}(k)) and slightly negative $\sigma_{xy}$ in the 60$^\circ$ NF configuration (Figures~\ref{fig:mixed-mechanical-60}(k)). Overall, the differences in strain and stress patterns between the pre-fractured NF configurations define distinct local stress states prior to fluid injection, which subsequently control HF initiation and propagation behavior. 

The largest discrepancies between numerical and experimental results occur near the NF interfaces, where steep gradients and stress concentrations develop. Minor differences may also arise from the extrapolation of DIC-derived displacement fields onto the XFEM mesh, particularly in regions of sharp displacement gradients at the NFs. 

\subsubsection{HF Deflection Controlled by Local Stress State} 

The stress redistribution established during mechanical loading governs HF propagation during subsequent fluid injection, producing deflection patterns that depend on the NF orientation relative to the far-field loading direction, $\sigma_{\max}$. In particular, the sign and spatial distribution of the shear stress and strain components, $\sigma_{xy}$ and $\varepsilon_{xy}$, developed during loading determine whether the local stress state promotes HF deflection toward or away from the NF.

\begin{figure}[]
  \centering
  \includegraphics[width=0.98\textwidth]{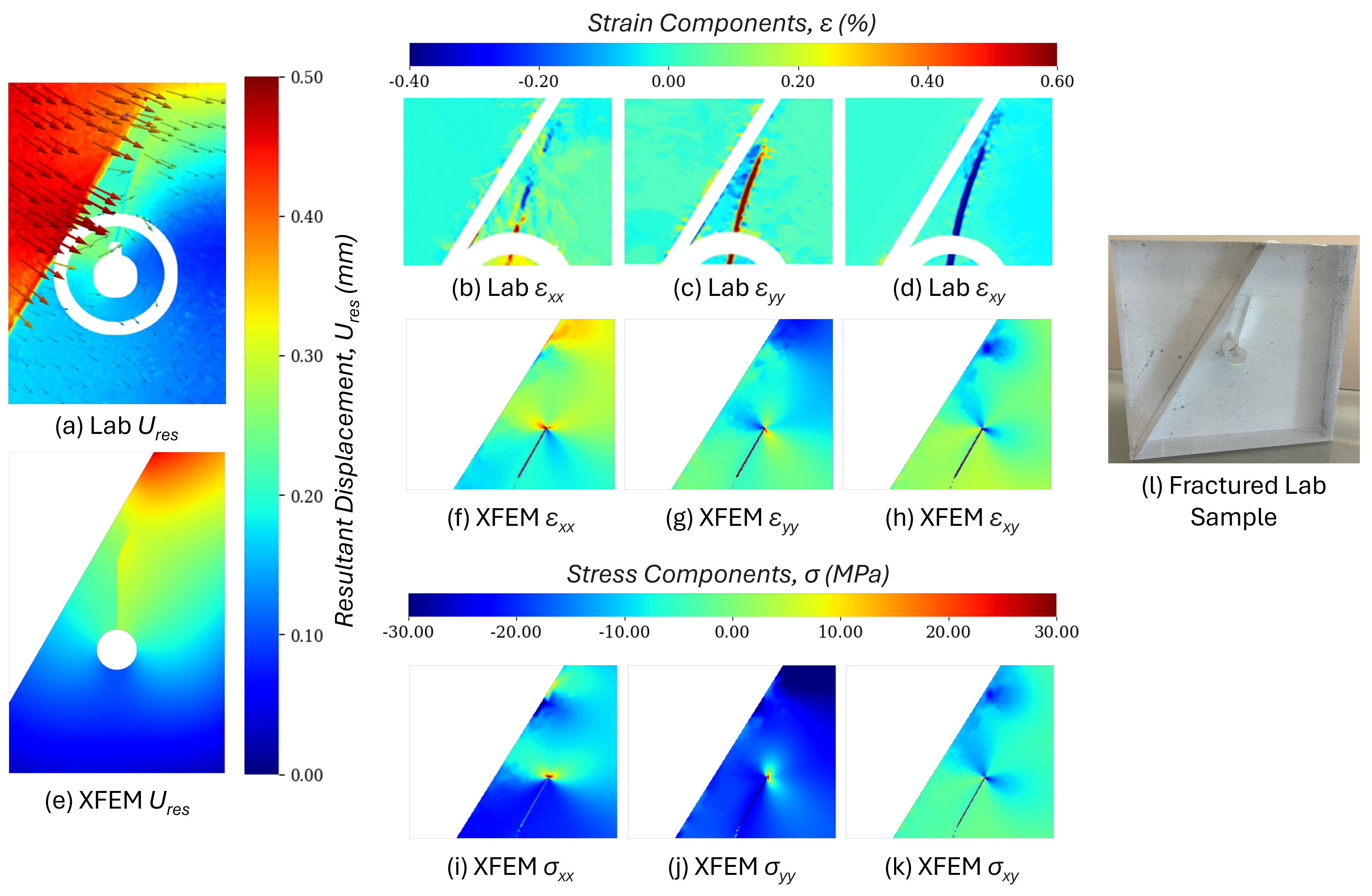}
  \caption{Displacement, strain, and stress fields for the 30$^\circ$ NF configuration PMMA specimen after hydraulic fracture. Experimental results obtained from DIC are shown in panels (a–d), and corresponding numerical results from XFEM are shown in panels (e–k). The top row shows DIC results: (a) resultant displacement in the  76~mm × 51~mm field of view and (b-d) horizontal, vertical, and shear strain fields in the region above the notch. Note: Borehole, notch, and O-ring regions are excluded from the DIC analyses. Middle and bottom rows are XFEM results: (e) resultant displacement in the 76~mm × 51~mm field of view, (f‒h) horizontal, vertical, and shear strains, and (i‒k) horizontal, vertical, and shear stresses in the region above the borehole. (l) A digital image of a hydraulically fractured Pre-fractured PMMA specimen.}
  \label{fig:HF-30}
\end{figure}

In the 30$^\circ$ NF configuration, the mechanically induced stress field is characterized by positive $\varepsilon_{xy}$ and $\sigma_{xy}$ (Figures~\ref{fig:HF-30}(d) and (h)), indicating a clockwise redistribution of shear relative to the far-field compressive loading. This shear redistribution is accompanied by increased normal compression along the NF and a comparatively tensile stress state on the right-hand side of the notch. As fluid injection proceeds, the HF encounters an asymmetric stress field in which elevated normal compression near the NF inhibits vertical propagation, while the opposite side favors tensile opening. The resulting normal-stress gradient, together with shear redistribution, deflects the HF away from the NF and toward the tensile corridor on the right.

Despite this pronounced deflection during propagation, the breakdown pressure for the 30$^\circ$ NF specimen remains comparable to that of the intact case (Figure~\ref{fig:pressure-evolution}(c)). This indicates that while the inclined NF strongly influences post-initiation fracture trajectory through shear-driven stress redistribution, it does not significantly relax the normal stress acting at the notch during fracture initiation for this configuration.

\begin{figure}[]
  \centering
  \includegraphics[width=0.98\textwidth]{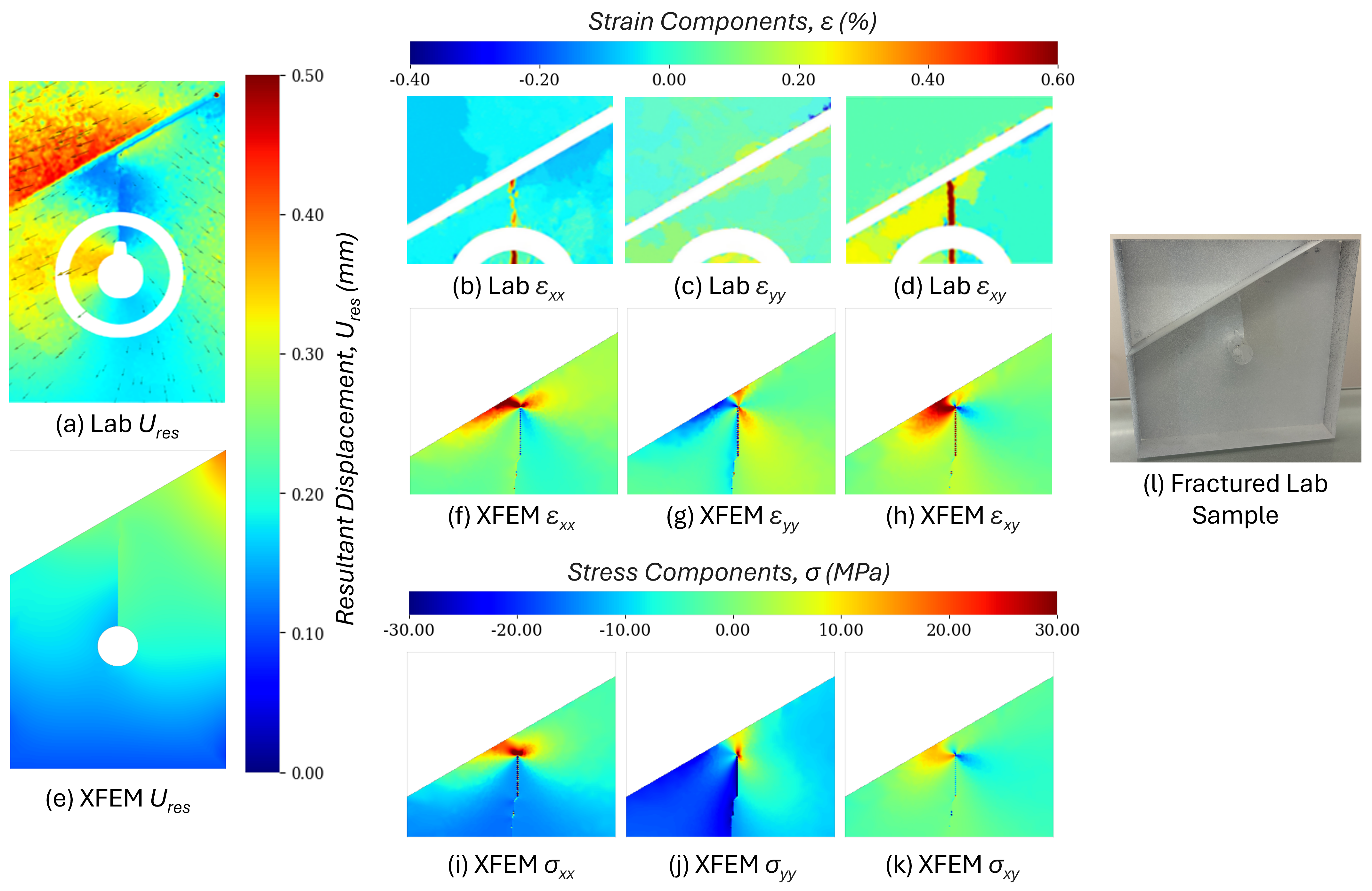}
  \caption{Displacement, strain, and stress fields for the 60$^\circ$ NF configuration PMMA specimen after hydraulic fracture. Experimental results obtained from DIC are shown in panels (a–d), and corresponding numerical results from XFEM are shown in panels (e–k). The top row shows DIC results: (a) resultant displacement in the 76~mm × 51~mm  field of view and (b-d) horizontal, vertical, and shear strain fields in the region above the notch. Note: Borehole, notch, and O-ring regions are excluded from the DIC analyses. Middle and bottom rows are XFEM results: (e) resultant displacement in the 76~mm × 51~mm  field of view, (f‒h) horizontal, vertical, and shear strains, and (i‒k) horizontal, vertical, and shear stresses in the region above the borehole. (l) A digital image of a hydraulically fractured pre-fractured PMMA specimen.}
  \label{fig:HF-60}
\end{figure}

In the 60$^\circ$ NF configuration, the mechanically induced stress field is characterized by negative $\varepsilon_{xy}$ and $\sigma_{xy}$ components (Figures~\ref{fig:HF-60}(d,h,k)), indicating an opposite sense of shear redistribution relative to the far-field loading direction compared to the 30$^\circ$ case. This shear redistribution reduces the effective normal stress acting at the notch, facilitating fracture initiation and explaining the lower breakdown pressure observed relative to the intact and 30$^\circ$ configurations (Figure~\ref{fig:pressure-evolution}(c)). As fluid injection proceeds, elevated tensile stresses develop along the inclined NF, creating a preferential pathway that attracts the propagating HF.

During propagation, increasing magnitudes of $\varepsilon_{xy}$ and $\sigma_{xy}$ (Figures~\ref{fig:HF-60}(d,h)) indicate an increasing shear contribution to the fracture process. The corresponding stress field further reveals compressive $\sigma_{xx}$ on the right side of the HF and tensile $\sigma_{xx}$ on the left (Figure~\ref{fig:HF-60}(i)), establishing a lateral gradient in normal stress that progressively draws the crack tip toward the NF. Together, these observations demonstrate the development of mixed-mode I–II fracture behavior driven by NF-induced shear redistribution and asymmetric normal stress conditions.


Table~\ref{tab:HF-NF-summary} provides a comparative summary of the observed $\varepsilon_{xy}$ and $\sigma_{xy}$ signs, corresponding HF curvature, and governing mechanisms for the intact, 30$^\circ$, and 60$^\circ$ NF specimens.


\begin{table}[]
\centering
\caption{Summary of experimental observations and inferred hydro-mechanical mechanisms for each configuration, highlighting the role of shear strain and stress components in controlling HF trajectory.}

\label{tab:HF-NF-summary}
\begin{tabular}{
    p{2 cm}  
    p{3 cm}  
    p{3.5 cm}  
    p{5 cm}  
}
\toprule
\textbf{Configuration} & 
\textbf{Shear Sign Relative to $\sigma_{\max}$} & 
\textbf{HF Curvature} & 
\textbf{Dominant Mechanism} \\
\midrule
\textbf{Intact} & 
$\approx 0$ & 
None, vertical & 
Pure tensile opening. \\

\textbf{30$^\circ$ NF} & 
Positive & 
Rightward, away from NF & 
Shear-induced compression barrier causing repulsion. \\

\textbf{60$^\circ$ NF} & 
Negative  & 
Leftward, toward NF & 
Shear-induced tensile corridor promoting attraction. \\
\bottomrule
\end{tabular}
\vspace{10pt}

\end{table}

\newpage
\section{Conclusions}
\label{sec:conclusions}
This study investigates how the presence of pre-existing NFs redistributes local stress components and governs HF paths in intact and pre-fractured PMMA specimens. By integrating full-field DIC measurements with poroelastic XFEM simulations, the coupled hydro-mechanical mechanisms controlling HF are resolved. The main conclusions are summarized as follows:

\begin{itemize}
    \item [-] Under uniform vertical compression and lateral confinement, the intact configuration exhibits a localized tensile zone above the borehole, accompanied by negligible shear strain and shear stress. This stress state promotes straight vertical HF aligned with propagation the far field maximum principal stress, corresponding to a purely Mode~I opening regime. This configuration establishes a baseline response in the absence of NF-induced stress perturbations.

    \item [-] Both the $30^\circ$ and $60^\circ$ NF configurations exhibit evident stress redistribution during mechanical loading, which generates asymmetric displacement fields with notable shear strain and shear stresses near the notch. These NF-induced stress redistributions alter the local stress state prior to fluid injection, creating distinct fracture conditions that cannot be predicted solely from far-field stress orientation or geometric criteria. As such, fracture trajectory cannot be reliably inferred from NF orientation alone, but instead requires resolving the evolving local stress state during loading and injection.

    \item [-] The sign of the shear components governs the direction of HF deflection relative to the NF. Positive shear in the $30^\circ$ NF configuration generates a localized compressive stress barrier adjacent to the NF, causing the HF to curve away from it. Conversely, negative shear in the $60^\circ$ NF configuration reduces the effective normal stress along the NF, creating a tensile pathway that attracts the HF toward the NF. This one-to-one correspondence establishes shear polarity as the primary mechanism controlling HF–NF interaction.

    \item [-] In both NF configurations, DIC and XFEM results demonstrated significant shear contributions to HF growth. The numerical simulations show that pure Mode I formulations cannot reproduce the experimentally observed curved trajectories, whereas mixed-mode (Mode I-II) energy release criteria are required for accurate prediction of the HF paths.


\end{itemize}


Future work will extend the DIC-based analysis to three-dimensional by integrating strain gauges and acoustic emission (AE) monitoring to quantify the effects of small but non-negligible lateral strains that are not fully captured under the near plane-strain assumption. 
\section{Acknowledgment}
The laboratory experiments conducted in this study were partially supported by the National Science Foundation through CAREER Award No. 2443533 and by the Department of Civil and Environmental Engineering at the University of Utah. The authors, Shahrzad Roshankhah, Shivesh Shandilaya, and Sunghyun Kim, also acknowledge Dr. Owen Kingstead for providing access to the VIC-2D software used for digital image analysis. The computational studies in this work were partially supported by the Sand Hazards and Opportunities for Resilience, Energy, and Sustainability (SHORES) Center, funded by Tamkeen under the NYU Abu Dhabi Research Institute Award CG013, and by the NYUAD Kawader Program. The authors, Mostafa Mobasher and Maithah Alaleeli, also acknowledge the NYUAD Center for Research Computing for providing computational resources, services, and technical support.

\newpage
\appendix
\section*{Appendix}
\numberwithin{equation}{section}
\section{Digital Image Correlation (DIC) Analysis}
\label{sec:DIC}
Full-field in-plane displacement and strain fields are obtained using two-dimensional digital image correlation (2D-DIC) applied to a speckle-patterned surface imaged in an undeformed reference state and in subsequent deformed states acquired during loading. A region-of-interest (ROI) mask is defined to restrict the analysis to the specimen area, excluding regions not intended for correlation. Further, analysis is performed using overlapping subsets defined by a specified subset size and step size. Larger subsets generally reduce displacement variability but may smooth localized deformation, whereas smaller subsets improve spatial resolution at the expense of increased noise sensitivity (Pan et al., \citeyear{pan2008study}; Reu et al., \citeyear{reu2015camera}). Imaging parameters, including lens focal length and camera stand-off distance, influence DIC accuracy by affecting optical distortion and sensitivity to local deformation (Jones et al., \citeyear{jones2018distortion}). Prior to full analysis, correlation-quality visualization is used to verify that the speckle texture within the ROI provides reliable tracking and spatially consistent correlation performance.

\subsection{Correlation criterion}
The accuracy of DIC measurements depends strongly on the chosen correlation formulation. In this study, deformation is quantified using the zero-normalized sum of squared differences (ZNSD) criterion, which is widely adopted due to its reduced sensitivity to illumination changes, camera response, and image noise \citep{pan2009two}. Correlation parameters (subset size, step, interpolation, and convergence settings) are selected to strike a balance between spatial resolution and measurement uncertainty. 

For a subset containing $N$ pixels, the ZNSD criterion is defined as:
\begin{equation}
C_{\mathrm{ZNSD}}
=
\sum_{i}\sum_{j}
\left(
\tilde{f}(x_i,y_j)
-
\tilde{g}(x'_i,y'_j)
\right)^2,
\label{eq:ZNSD}
\end{equation}
where the normalized grayscale intensities are:
\begin{equation}
\tilde{f}(x_i,y_j)=\frac{f(x_i,y_j)-f_m}{\Delta f},
\qquad
\tilde{g}(x'_i,y'_j)=\frac{g(x'_i,y'_j)-g_m}{\Delta g}.
\end{equation}
Here, $f(x_i,y_j)$ and $g(x'_i,y'_j)$ denote grayscale intensities in the reference and deformed images, respectively; $f_m$ and $g_m$ are subset mean intensities; and $\Delta f$ and $\Delta g$ are the corresponding standard deviations. Sub-pixel displacements are obtained by minimizing $C_{\mathrm{ZNSD}}$ using an iterative nonlinear least-squares procedure, consistent with VIC-2D documentation.

\subsection{Displacement and strain computation}
Following the correlation, in-plane displacement components ($u,v$) are obtained on a regular grid. Strain fields are computed within VIC-2D from the displacement grid using a Lagrangian strain formulation,
\begin{align}
\varepsilon_{xx} &= \frac{\partial u}{\partial x}, \\
\varepsilon_{yy} &= \frac{\partial v}{\partial y}, \\
\varepsilon_{xy} &= \frac{1}{2}
\left(
\frac{\partial u}{\partial y}
+
\frac{\partial v}{\partial x}
\right),
\end{align}
with displacement gradients evaluated over a strain radius of 15 pixels. The strain radius defines the neighborhood used for gradient estimation and controls the trade-off between numerical stability and spatial resolution, in accordance with VIC-2D strain computation guidelines.

\subsection{Post-processing and denoising}
To further reduce noise in DIC-derived scalar fields, MATLAB-based post-processing is performed using a non-local means (NLM) denoising algorithm (Buades et al., \citeyear{buades2006staircasing}; Buades et al., \citeyear{buades2008nonlocal}; Rubino et al., \citeyear{rubino2015static}; Tal et al., \citeyear{tal2019enhanced}). For a pixel at location $(i,j)$, a reference patch $W_1$ of size $(2f+1)\times(2f+1)$ is compared with candidate patches $W_2$ within a search window of size $(2t+1)\times(2t+1)$, where $f$ denotes the patch radius and $t$ the search-window radius. Symmetric padding is applied prior to patch extraction to mitigate any boundary effects.

Patch dissimilarity is computed as a kernel-weighted sum of squared differences,
\begin{equation}
d = \sum \sum \left( \mathbf{K} \odot (W_1 - W_2)^2 \right),
\end{equation}
where $\mathbf{K}$ is a normalized weighting kernel. Similarity weights are assigned using an exponential decay law,
\begin{equation}
\omega = \exp\left(-\frac{d}{h^2}\right),
\end{equation}
with $h$ controlling the filtering strength. The filtered value is obtained as the normalized weighted average of candidate values within the search window, with explicit inclusion of the maximum weight to enhance numerical stability. The parameters $(f,t,h)$ therefore define the patch size, search neighborhood, and degree of smoothing, respectively.

\subsection{Vector-field visualization}
Vector-field visualization is performed in ParaView using a consistent and reproducible workflow applied to CSV-formatted outputs. Displacement components are assembled into vector form and rendered using arrow glyphs to inspect displacement magnitude and direction across frames. This workflow is applied consistently to generate all displacement vector visualizations presented in this study.

\section{Derivation of Governing Equations}
\label{Appendix:Num}

\subsection{Poroelastic Deformation}

The deformation of a saturated porous medium under mechanical and hydraulic loads is described by the theory of linear poroelasticity (Biot, \citeyear{biot1941}), which couples the elastic response of the solid skeleton with the fluid pressure field of the interstitial fluid. In the absence of body forces, the static mechanical equilibrium is expressed through the local balance of linear momentum:
\begin{equation}
\nabla \cdot \boldsymbol{\sigma} = \mathbf{0}
\end{equation}
where \( \boldsymbol{\sigma} \) is the total Cauchy stress tensor (Pa). The total stress is related to the effective stress \( \boldsymbol{\sigma}' \) (Pa) and the pore pressure \( p \) (Pa) through Biot’s relation (Biot, \citeyear{biot1941}):
\begin{equation}
\boldsymbol{\sigma} = \boldsymbol{\sigma}' - \alpha p \mathbf{I}
\end{equation}
where \( \alpha \) is the Biot coefficient (\( 0 \leq \alpha \leq 1 \)) and \( \mathbf{I} \) is the identity tensor. Assuming small deformations, the effective stress is dictated by the isotropic linear elastic constitutive law (Terzaghi, \citeyear{tr1943}):
\begin{equation}
\boldsymbol{\sigma}' = 2G \boldsymbol{\varepsilon} + \lambda \, \text{tr}(\boldsymbol{\varepsilon}) \mathbf{I}
\end{equation}
where \( G \) and \( \lambda \) are Lamé’s parameters (Pa), and \( \boldsymbol{\varepsilon} \) is the infinitesimal strain tensor defined as:
\begin{equation}
\boldsymbol{\varepsilon} = \frac{1}{2} \left( \nabla \mathbf{u} + (\nabla \mathbf{u})^\mathrm{T} \right)
\end{equation}
with \( \mathbf{u} \) being the displacement vector field (m). This formulation ensures that deformation is driven not only by externally applied loads but also by changes in the internal pore pressure.

\subsection{Fluid Flow in the Matrix}

The motion of fluid within the porous matrix is governed by the continuity equation, which enforces local conservation of mass for a slightly compressible fluid in a deforming porous medium \citep{biot1941}:
\begin{equation}
\label{Eq:continuity}
\frac{1}{M} \frac{\partial p}{\partial t} + \alpha \frac{\partial \varepsilon_v}{\partial t} + \nabla \cdot \mathbf{v}_p = 0
\end{equation}
where \( p \) is the pore pressure (Pa), \( M \) is the Biot modulus (Pa), \( \varepsilon_v = \text{tr}(\boldsymbol{\varepsilon}) \) is the volumetric strain of the solid matrix, and \( \mathbf{v}_p \) is the Darcy seepage velocity (m/s). The coupling term \( \alpha \, \partial \varepsilon_v / \partial t \) accounts for changes in pore volume induced by deformation of the solid skeleton.

Fluid flow within the matrix follows Darcy’s law \citep{whitaker1986flow}, which relates the seepage velocity to the pressure gradient:
\begin{equation}
\mathbf{v}_p = -\frac{k}{\mu} \nabla p = -\frac{\bar{k}}{\gamma} \nabla p
\end{equation}
where \( k \) is the intrinsic permeability (m$^2$), \( \mu \) is the dynamic viscosity of the pore fluid (Pa$\cdot$s), \( \bar{k} \) is the hydraulic conductivity (m/s), and \( \gamma \) is the specific weight of the fluid (N/m$^3$).

Substituting Darcy’s law into Eq.~(\ref{Eq:continuity}) yields the pressure diffusion equation:
\begin{equation}
\frac{1}{M} \frac{\partial p}{\partial t} + \alpha \frac{\partial \varepsilon_v}{\partial t}
= \nabla \cdot \left( \frac{k}{\mu} \nabla p \right),
\end{equation}
which governs the evolution of the pore-pressure field due to the combined effects of deformation-induced pore-volume changes and pressure-driven fluid transport.

\subsection{Fracturing Fluid Flow}

Within the fracture, the injected fluid flows between two separating fracture faces, as illustrated in Figure~\ref{fig:fracturing_flow}. Assuming laminar flow of an incompressible, Newtonian fluid, lubrication theory is applicable \citep{reynolds1886iv}. Under these assumptions, the fracture is idealized as a narrow, smooth-walled channel with aperture much smaller than its length.

Integrating the velocity profile across the fracture aperture yields the Poiseuille relation for the volumetric flow rate per unit fracture width:
\begin{equation}
\label{Eq:Poiseuille}
q(s) = -\frac{w^3(s)}{12\mu_f} \frac{\partial p_f}{\partial s},
\end{equation}
where \( q(s) \) is the volumetric flow rate per unit fracture width (m$^3$/s/m), \( s \) is the curvilinear coordinate along the fracture, \( w(s) \) is the local fracture aperture (m), \( \mu_f \) is the dynamic viscosity of the fracturing fluid (Pa$\cdot$s), and \( p_f \) is the fracture fluid pressure (Pa). The cubic dependence on aperture highlights the strong coupling between fracture opening and fluid transport.

Conservation of mass for the fluid within the fracture is expressed as:
\begin{equation}
\label{Eq:fracture_continuity}
\frac{\partial q}{\partial s} + q_{\text{leak}} = Q_{\text{inj}}(s),
\end{equation}
where \( Q_{\text{inj}}(s) \) is the local fluid injection rate per unit fracture width, and \( q_{\text{leak}} \) represents the total leak-off flux from the fracture into the surrounding porous matrix.

Leak-off across the fracture surfaces is modeled as:
\begin{equation}
q_{\text{leak}} = C_t \left( p_f - p_{\text{top}} \right) + C_b \left( p_f - p_{\text{bot}} \right),
\end{equation}
where \( C_t \) and \( C_b \) are the leak-off coefficients (m$^3\cdot$Pa$^{-1}$/s) for the top and bottom fracture faces, respectively, and \( p_{\text{top}} \) and \( p_{\text{bot}} \) are the adjacent matrix pore pressures (Pa).

Substituting Eq.~(\ref{Eq:Poiseuille}) into Eq.~(\ref{Eq:fracture_continuity}) yields the nonlinear diffusion equation governing fracture fluid pressure:
\begin{equation}
\frac{\partial}{\partial s} 
\left(
-\frac{w^3}{12\mu_f} \frac{\partial p_f}{\partial s}
\right)
+ C_t \left( p_f - p_{\text{top}} \right)
+ C_b \left( p_f - p_{\text{bot}} \right)
= Q_{\text{inj}}(s),
\end{equation}
which couples fracture aperture evolution, fluid leak-off, and pressure redistribution within the fracture.

\subsection{Fracture Initiation and Propagation}

Fracture initiation and growth can be conceptualized as a transition between two limiting states:  
(i) an initially intact medium in which displacements are continuous and tractions are sustained across all material interfaces, and  
(ii) a fully fractured state characterized by a displacement discontinuity with vanishing normal tractions across the fracture surface.  
This transition is commonly described using cohesive fracture mechanics through traction--separation laws, in which interface strength degrades progressively with increasing displacement jump \citep{Barenblatt1962,Ortiz1999}.

In this study, fracture initiation and propagation are modeled using the extended finite element method (XFEM). A key advantage of XFEM is that it enriches the displacement and pore-pressure fields within standard continuum elements, allowing fractures to nucleate and propagate dynamically without predefined paths or remeshing \citep{Abaqus2016}. This capability is particularly well suited for simulating hydraulic fracture growth in complex geometries and heterogeneous stress fields.

Fracture growth is represented using a linear softening traction--separation law:
\begin{equation}
T_n(\delta_n) =
\begin{cases}
K_0 \,\delta_n, & 0 \leq \delta_n \leq \delta_0, \\[4pt]
\sigma_c \left( 1 - \dfrac{\delta_n - \delta_0}{\delta_c - \delta_0} \right), & \delta_0 < \delta_n \leq \delta_c, \\[6pt]
0, & \delta_n > \delta_c,
\end{cases}
\label{eq:TSL}
\end{equation}
where $T_n$ is the normal traction (Pa), $\delta_n$ is the normal separation (m), $K_0$ is the initial interface stiffness (Pa/m), $\sigma_c$ is the cohesive strength (Pa), $\delta_0 = \sigma_c/K_0$ is the separation at damage initiation (m), and $\delta_c$ is the critical separation at complete failure (m). The traction initially increases linearly with opening displacement until the cohesive strength $\sigma_c$ is reached, after which it decreases linearly to zero at $\delta_c$, corresponding to full decohesion.

As illustrated in Figure~\ref{fig:traction}, the area under the traction--separation curve defines the fracture energy $G_c$ (J/m$^2$), which represents the energy required to create new fracture surfaces. Damage initiation occurs when the maximum principal stress exceeds $\sigma_c$, after which interface stiffness progressively degrades and a displacement discontinuity develops. Once activated, the fracture becomes hydraulically conductive, and the fracturing fluid pressure $p_f$, governed by Eq.~(\ref{Eq:Poiseuille}), is applied as a boundary load across the evolving fracture faces. This hydro-mechanical coupling drives continued fracture opening and propagation, enabling fully coupled simulation of fluid-driven fracture growth in poroelastic media \citep{Zielonka2014}.

\printcredits
\newpage
\bibliographystyle{cas-model2-names}

\bibliography{cas-refs}



\end{document}